\documentclass{article} 

\usepackage{cite}
\usepackage{graphicx}

\begin{document}
\begin{center}

{\large\sc\bf Parameterizing the flattening of galaxies rotation curves on an expanding locally anisotropic background\\}

\vspace{1 truecm}

\vspace{0.8 truecm}
{\bf P. Castelo Ferreira}\footnote{\texttt{pedro.castelo.ferreira@gmail.com}}\\[10mm] {\small Center for Rapid and Sustainable Product Development\\ Polytechnic Institute of Leiria\\[2mm]}
\ \\[20mm]
{\sc\bf Abstract}
\end{center}

In this paper are discussed possible many body generalizations of the expanding locally anisotropic metric ansatz with respect to approximately Newtonian gravitational systems. This ansatz consistently describes local point-like matter distributions on the expanding Universe also allowing for a covariant parameterization of gravitational interactions at intermediate length scales.

As an example of applicability it is modeled a disk galaxy model matching the physical parameters of the galaxy UGC2885 and it is shown that, by fine-tuning the metric functional parameter, the flattening of the galaxy rotation curve is fully parameterized by this metric. In addition it is numerically computed the mass-energy density corrections due to the expanding anisotropic background and explicitly shown that although there are negative contributions within the galaxy plane the total mass-energy density is strictly positive both at the galaxy plane and outside the galaxy plane. As the functional parameter for this metric is a dimensionless exponential factor is required a floating point precision of $250$ significant digits for root finding routines and $200$ significant digits to evaluate the effective mass-energy density rendering a final precision of the results presented above double precision ($16$ significant digits).

It is further shown that these results are consistent with the interpretation of the gravitational corrections as due to Dark Matter, in particular constituting a novel heuristic local parameterization for the Dark Matter distribution within the galaxy plane consistent with both local scale and cosmological scale physical laws which is useful to further investigate the local properties of Dark Matter.

%\pacs{95.35.+d}{Dark Matter}
%\pacs{98.80.Jk}{Mathematical and relativistic aspects of cosmology}
%\pacs{04.20.-q}{Classical General Relativity}

\thispagestyle{empty}
\newpage
\section{Introduction}

The main objective of this paper is to generalize in the context of many body gravitational systems the recently suggested expanding locally anisotropic metric ansatz~\cite{PLB,e-print} as well as to show that such ansatz allows to parameterize corrections to the gravitational acceleration within intermediate gravitational scales which is commonly attributed to Dark Matter. This ansatz describes point-like local matter distributions in an expanding background, the Universe~\cite{Hubble},  As an example, it is shown that for a specific fine-tuning of the functional parameter of this metric the observed flattening of galaxies rotation curves with respect to the predicted classical profiles, when considering only the Newtonian acceleration due to baryonic matter within galaxies~\cite{v_gal1,mass_1}, is fully described such that this gravitational background ansatz can be interpreted as a covariant parameterization of the gravitational corrections attributed to Cold Dark Matter~\cite{DM}. Hence this study allows for a heuristic parameterization of Dark Matter within the context of General Relativity maintaining as the asymptotic limit the cosmological standard model which is useful to investigate the local properties of Dark Matter. Assuming a thin exponential disk approximation we apply such parameterization to a simplified lattice model matching the characteristics of the galaxy UGC2885~\cite{v_gal2,v_gal3,v_gal3a,v_gal4} and numerically estimate the mass-energy density required to match the observed galaxy rotation curve.

The ELA metric ansatz was derived in~\cite{PLB} as a generalization of previous metric solutions and ansatze~\cite{McV,McV2} describing local matter distributions in the expanding Universe~\cite{Hubble} and noting that, although global space-time isotropy is mandatory, local matter distributions generate local anisotropy~\cite{anisotropy,wmap}. Specifically the infinitesimal line element for this ansatz is 
\begin{equation}
\begin{array}{rcl}
ds_{ELA}^2&=&\displaystyle(1-U)(cdt)^2\\[5mm]
&&\displaystyle-\frac{1}{1-U}\left(dr_1-\frac{H\,r_1}{c}(1-U)^{\frac{\alpha}{2}+\frac{1}{2}}(cdt)\right)^2-d\Omega^2\\[5mm]
&=&\displaystyle\left(1-U-\left(\frac{H\,r}{c}\right)^2\left(1-U\right)^\alpha\right)(cdt)^2\\[6mm]
&&\displaystyle + \frac{H\,r_1}{c}(1-U)^{\frac{\alpha}{2}-\frac{1}{2}}dr_1\,(cdt)-\frac{1}{1-U}\,dr_1^2-d\Omega^2
\end{array}
\label{metric_1}
\end{equation}
where $U=2GM/(c^2r_1)$ is the Schwarzschild (SC) gravitational potential, $H=\dot{a}/a$ is the time dependent Hubble rate defined as the
rate of variation of the Universe scale factor $a$, $c$ is the speed of light in vacuum,
$M$ is the SC gravitational mass, $G$ is the Gravitational constant, $d\Omega^2=r_1^2(d\theta^2+\sin^2\theta d\varphi)$ is the solid angle
line-element and we are employing spherical coordinates $(t,r_1,\varphi,\theta)$ for which the radial coordinate is an area radius
(the area of a sphere is $A=4\pi\,r_1^2$) such that the integration measure is independent of the Universe scale factor, $\sqrt{-g}=r_1^2\sin(\theta)$. For $\alpha = 0$ it is recovered the McVittie metric~\cite{McV} and recently it was shown that for $\alpha=-1$ it is recovered the Thakurta metric~\cite{Thakurta}. The ELA metric~(\ref{metric_1}) interpolates between the SC metric~\cite{Schwarzschild} for small radii values (near the SC event horizon at the SC radius $r_1\sim r_{1.SC}=2GM/c^2$) and the Robertson-Walker metric~\cite{RW} for large radii values ($r_1\to \infty$). The shift function depends on a functional parameter $\alpha$ which, at the SC event horizon must be greater or equal to $\alpha(r_{1.SC})\geq 3$ to prevent this horizon from being a space-time singularity, at the origin must have at least a leading divergence by $\alpha(r_1\to 0)\sim -1/r_1$ to ensure that the SC mass pole value coincides with the SC mass $M$ and, at spatial infinity, it must be finite such that the shift function asymptotically vanishes and the RW metric is recovered~\cite{PLB}.

As for the mass of the extended gravitational background encoded in the ELA metric in excess to the mass of the cosmological
background described by the RW metric it was shown in~\cite{Pioneer} to be
\begin{equation}
\begin{array}{rcl}
M_{\alpha}&=&\displaystyle\lim_{R_1\to +\infty} 4\pi\int_0^{R_1}r_1^2\left(\rho_{(\alpha)}-\rho_{RW}\right)dr_1\\
&=&\displaystyle \lim_{R_1\to +\infty}\frac{H^2}{4G}\left(\left(1-\frac{2GM}{c^2R_1}\right)^{\alpha(R_1)}-1\right)R_1^3\ .
\end{array}
\label{M_alpha}
\end{equation}
In this expression $\rho_{(\alpha)}$ is the (extended) mass-energy density contribution due to the expanding locally anisotropic metric in excess of the SC mass pole of value 
$M$ and $\rho_{RW}=3H^2/(8\pi G)$ is the cosmological background mass-energy density. The value for this mass is finite either when it is considered a radial cut-off for $\alpha$ above which this functional parameter is
null such that the isotropic McVittie metric~\cite{McV} is exactly recovered, or when the functional parameter is asymptotically null
at spatial infinity being asymptotically proportional to $\sim\frac{1}{r^n}$ with $n\geq 2$. Hence either of these limits for $\alpha$ must be considered
to ensure a finite mass contribution due to the expanding anisotropic background.

In addition, in between the two well established asymptotic limits (the SC metric and RW metric), there are no specific bounds on the functional parameter $\alpha$ such that it allows for a covariant parameterization of unmodeled observable effects at intermediate length scales. Also we recall that so far no direct physical interpretation for the functional parameter $\alpha$ exist, here the ELA metric ansatz is interpreted at most as a covariant parameterization of gravitational interactions corrections compatible with the well established asymptotic background solutions, the local SC metric and the RW metric. In this work we explore the application of such parameterization to describe the dynamics of many body gravitational systems, namely the flattening of rotation curves for galaxies, which indicates that such parameterization may be physically interpreted as Cold Dark Matter. Such parameterization has also been applied to the Solar System dynamics~\cite{Pioneer}, in particular allowing for a parameterization of the variation of the Astronomical Unit~\cite{AU}.

As a final remark aiming at allowing to independent reproduction of the results computed in this paper, let us note that as the functional parameter is explicitly written in the metric ansatz as a dimensionless exponential factor to obtain numerical meaningful results it is required to set the working precision for root find routines to at least $250$ significant digits and to evaluate the mass-energy density to at least $200$ significant digits. Hence the final output results presented in this paper are computed with double precision of $16$ significant digits. However it is relevant to stress that as the fundamental constants and galactic rotation curves experimental data have a much lower precision (typically below $8$ significant digits), the results discussed next are at most physically meaningful to the same precision of experimental data.

\section{Ansatze for many body systems}

With the objective of generalizing the ansatz~(\ref{metric_1}) to many body gravitational systems let us note that, generally, the above ansatz for the ELA metric~(\ref{metric_1}) is obtained from any given metric $g_{\mu\nu}$ describing the gravitational background of local matter distributions by considering the radial shift function $N^r_{ELA}=-H\,r_1\,(\sqrt{g_{00}})^{\alpha+1}/c$. Hence to generalize the ELA metric ansatz to many body backgrounds it is first require to define the many body SC metric background. However no explicit analytical solution for a metric describing such background is known. The standard approach is to consider a perturbative post-Newtonian solution to a specific order in the gravitational field $U$~\cite{PN}. Here we will consider the Newtonian approximation which is commonly employed in macroscopical galaxy models~\cite{disk} (see also~\cite{Peebles}), hence we define the local Newtonian gravitational background for N massive bodies to first order in the gravitational field
\begin{equation}
\begin{array}{rcl}
ds_{SC.N}^2&=&\displaystyle\left(1-U_N\right)(cdt)^2-dr^2-d\Omega^2\ ,\\[5mm]
U_N&=&\displaystyle\sum_{n=1}^N\,U_n\ \ \ ,\ \ \ U_n\,=\,\frac{2GM_n}{c^2|\mathbf{r_1-r_{1.n}}|}\ ,
\end{array}
\label{SC_N}
\end{equation}
where $U_N$ is the total SC gravitational potential being $U_n$ the SC gravitational potential for each of the many bodies. The distance to each body position is given as usual by the Euclidean distance between the point at which the metric is being evaluated ($\mathrm{r_1}$) and the position of each massive point-like body ($\mathbf{r_{1.n}}$). We recall that for simulations within the Solar System the gravitational field is at most of order $U<10^{-8}$ and a second order expansion on the gravitational field is commonly employed, hence being obtained an accuracy on the metric definition of order $\sim 10^{-16}$ which is enough to match the current experimental accuracy for observational data~\cite{Solar}. In this work we will consider a galaxy model based in an analytical thin disk approximation for which the SC gravitational field is typically of order $U_n < 10^{-7}$ such that the second order corrections would be at most of order $\sim 10^{-14}$. Hence the first order expansion considered for $g_{00.N}$ and $g_{rr.N}$ is accurate to order $\sim 10^{-7}$ being enough for exemplification purposes and to obtain a macroscopical estimative for the functional parameter of the ELA metric ansatz. A more detailed model would require a higher order expansion on the gravitational field.

Given the many body perturbative background~(\ref{SC_N}) it is straight forward to generalize the ELA metric ansatz by considering
the definition of the shift function with respect to the total gravitational potential 
\begin{equation}
\begin{array}{rcl}
N^r_{ELA.\mathrm{eff}}&=&\displaystyle -\frac{H\,r}{c}(1-U_N)^{\frac{\alpha_{\mathrm{eff}}}{2}+\frac{1}{2}}\\[5mm]
ds_{ELA.\mathrm{eff}}^2&=&\left(1-U_N\right)(cdt)^2-d\Omega^2\\[4mm]
&&\displaystyle-\left(dr_1+N^r_{ELA.\mathrm{eff}}(cdt)\right)^2\ .\label{metric_eff}
\end{array}
\label{metric_ELA_eff}
\end{equation}
This metric has the same properties of the ELA metric~(\ref{metric_1}), in particular at the event horizon
for each of the $n$ bodies there is no space-time singularity. We remark that the functional parameter
$\alpha_{\mathrm{eff}}$ parameterizes a collective gravitational correction to the SC metric due to the many bodies
in the gravitational system such that a direct fit to the galaxy rotation curve can be carried allowing to map this
functional parameter within the galaxy plane, hence to compute the mass-energy density and anisotropic pressures within
the galaxy.

To further compute the mass-energy density and anisotropic pressures outside the galaxy plane, it is
required to estimate the functional parameter $\alpha_{\mathrm{eff}}$ outside the galaxy plane. To achieve such estimate
it must be defined each individual body contribution in the gravitational system to the metric shift function. Also aiming
both at describing gravitational systems by more detailed models including the individual known massive bodies~\cite{AU}, as well
as to achieve a physical interpretation for the functional parameter of the ELA metric, it is desirable to have a functional parameter
for each individual body in any given gravitational system. Hence, next, we are mapping the shift function $N^r_{ELA.\mathrm{eff}}$ into
an equivalent product of the many body contributions defined with
respect to the gravitational fields of each individual body in the gravitational system. To achieve such factorization let us note
that the event horizons for each of the many body is no longer a spherical surface, instead the Schwarzschild surface
for each of these $n$ bodies is defined by the 2-dimensional solution to the equation $U_N=1$ in the neighborhood of each of the point-like massive bodies.
Specifically, for a given body $n$, the event horizon is the solution of the following equation
\begin{equation}
\frac{2GM_{n}}{c^2|\mathrm{r_{1.n}-r_{1.SC.n}}|}+\sum_{i\neq n}\frac{2GM_{i}}{c^2|\mathrm{r_{1.i}}-\mathrm{r_{1.SC.n}}|}\,=\,1\ .
\end{equation}
The solution of the radial coordinate $r_{1.SC.n}$ for each of the $N$ event horizons can be parameterized by the angular coordinates $\theta$
and $\varphi$ such that, for each of the $n$ bodies, this equation can be expressed in terms of a modified gravitational potential as $\tilde{U}_n=1$. Specifically, this modified gravitational potential, is defined as
\begin{equation}
\begin{array}{rcl}
\tilde{U}_n&=&\displaystyle \frac{2G\tilde{M}_n(\theta,\varphi)}{c^2|\mathrm{r_{1.n}-r_1}|}\ ,\\[5mm]
\tilde{M}_n(\theta,\varphi)&=&\displaystyle M_n+|\mathrm{r_{1.n}-r_{1.SC.n}(\theta,\varphi)}|\sum_{i\neq n}\frac{M_i}{|\mathrm{r_{1.i}-r_{1.SC.n}(\theta,\varphi)}|}\ .
\end{array}
\label{modified_U}
\end{equation}
For gravitational systems for which the values of the Schwarzschild radius is much smaller than the distance between bodies, $|\mathrm{r_{1.n}-r_{1.SC.n}(\theta,\varphi)}|\ll|\mathrm{r_{1.i}-r_{j.n}(\theta,\varphi)}|$, this modified gravitational potential
approximately matches the SC gravitational potential $\tilde{U}_n\approx U_n$.

Once we have defined these modified potentials, the shift function for the ELA metric can be factorized into a product of exponentials
$(1-\tilde{U}_n)$ which exactly vanish at the event horizon corresponding to each of the many body in the system and asymptotically converge
to unity at spatial infinity, hence having the same properties of the original single body ELA metric. Specifically we obtain the
ansatz
\begin{equation}
\begin{array}{rcl}
N^r_{ELA.N}&=&\displaystyle -\frac{H\,r}{c}\prod_{n=1}^N(1-\tilde{U}_n)^{\frac{\alpha_n}{2}+\frac{1}{2}}\\[5mm]
ds_{ELA.N}^2&=&\left(1-U_N\right)(cdt)^2-d\Omega^2\\[4mm]
&&\displaystyle-\left(dr_1+N^r_{ELA.N}(cdt)\right)^2\ ,\label{metric_N}\\[7mm]
\end{array}
\label{metric_ELA_N}
\end{equation}
The map between the functional parameters of metrics~(\ref{metric_N}) and~(\ref{metric_eff}) is straight forwardly obtained to be
\begin{equation}
\alpha_{\mathrm{eff}}=-1+\frac{\displaystyle\sum_{n=1}^N(\alpha_n+1)\ln(1-\tilde{U}_n)}{\displaystyle\ln\left(1-U_N\right)}\ .
\label{map}
\end{equation}
With respect to the gravitational interactions corrections, these two parameters have a distinct meaning. While the functional parameter $\alpha_n$ describes the gravitational interactions due to each body $n$ in the remaining $N-1$ bodies, the functional parameter $\alpha_{\mathrm{eff}}$ describes the local effect near each body $n$ due to the gravitational interactions of the remaining $N-1$ bodies.

Most astrophysical systems, such as galaxies, lye approximately on a 2-dimensional plane. Hence the physical laws derived
from direct observational data of a given astrophysical system are valid within such plane. The analysis of physical laws
in the neighborhood of the system is usually inferred from known gravitational laws as well as by analyzing  indirect
observational data such as the gravitational lens effect. Following this discussion, from the map~(\ref{map}), known
a profile for the functional parameter $\alpha_{\mathrm{eff}}$ it is possible to derive a profile for each of the functional
parameters $\alpha_n$ aiming at obtaining a estimation for the gravitational corrections due to the ELA metric outside of the plane
of the astrophysical system.

For a given gravitational system we are considering the following simplification assumptions: 
\begin{itemize}
\item the system is approximately planar allowing for a circular symmetric planar model;
\item each of the functional parameters $\alpha_n$ is circular symmetric with respect to the center of mass $r_{1.n}$ of each body $M_n$
in the plane of the system;
\end{itemize}
Given these assumptions the map~(\ref{map}) allows to obtain a scaling law for the functional parameter $\alpha_n$ with respect to each
body mass $M_n$. Let us consider $N_M$ bodies, each with a mass $M_0$, hence with total mass $M=N_M\,M_0$. If we consider these $N_M$ bodies to be infinitesimally close to each other at a distance $r_{1.0}$ from a given test mass such that $U_{i_N}=U_0=2GM_0/(c^2\,r_{1.0})$ and
$\tilde{U}_{i_N}=NU_0$ (for $i_N=1,\ldots,N$), the effective functional parameter $\alpha_{N_M}$ is, according to map~(\ref{map}), $\alpha_{N_M}=-1+N(\alpha_0+1)$. Further noting that $N_M=M/M_0$ and that the $N_M$ bodies must be consistently described by one
single body of mass $M$ we obtain the scaling law for the functional parameters $\alpha_n$ with respect to some reference mass $M_0$ to be
\begin{equation}
\alpha_{n}=-1+\frac{M_n}{M_0}(\alpha_0+1)\ .
\label{scaling_M}
\end{equation}
In particular, assuming that this scaling law is circular symmetric, allows to obtain a reference profile for the functional parameter $\alpha_0$
on the planar system from a known profile for $\alpha_{\mathrm{eff}}$, also on the planar system. Further assuming either approximately spherical symmetry for each functional parameter $\alpha_n$ or by parameterizing its anisotropy along the orthogonal direction to the planar system the functional parameter $\alpha_{\mathrm{eff}}$ can be computed outside the planar system. For a given many body system, we will explore this relation to estimate a $3D$ map for $\alpha_{\mathrm{eff}}$. We also remark that this scaling
law implies relatively large values of the parameter $\alpha_M$ for large masses as it scales linearly with the mass.

In addition, assuming the above scaling relation for each of the many bodies with masses $M_n$, in the limit of large radii the map~(\ref{map})
is approximately given by
\begin{equation}
\begin{array}{rcl}
\alpha_{\mathrm{eff}}(r_1\sim+\infty)&=&\displaystyle-1+\sum_{n=1}^N\frac{M_{n}}{M_0}\frac{\tilde{M}_n(\theta,\varphi)}{M_N}\,\left(\alpha_0(r_1\sim+\infty)+1\right)\\[5mm]
&\approx&\displaystyle -1+\sum_{n=1}^N\frac{M_{n}^2}{M_0\,M_N}\left(\alpha_0(r_1\sim+\infty)+1\right)\ ,\\[5mm]
M_N&=&\displaystyle \sum_{n=1}^N M_{n}\ ,
\end{array}
\label{alpha_infty_M}
\end{equation}
where $M_N$ is the total mass of the gravitational system, the masses $\tilde{M}_n(\theta,\varphi)$ are defined in~(\ref{modified_U}) and the
last expression for $\alpha_{\mathrm{eff}}$ is obtained by considering the approximation $\tilde{M}_n\approx M_n$. Noting that far from the
astrophysical body the gravitational potential approximately equals the gravitational potential of a single point-like massive body with the
total mass of the N-body astrophysical system $U_N\approx 2G\,M_N/(c^2\,r_1)$ such that the mass contribution due to the extended anisotropic
gravitational background is expressed by equation~(\ref{M_alpha}) with $M=M_N$ and further recalling that, as discussed in the introduction, to
ensure a finite mass contribution the metric functional parameter must either be asymptotically null at spatial infinity or strictly null
above a finite radial cut-off, we conclude that the above asymptotic expression for $\alpha_{\mathrm{eff}}(r_1\sim+\infty)$ must be null. Hence,
allowing for a generic value for the reference mass $M_0$, we obtain
\begin{equation}
\alpha_{\mathrm{eff}}(r_1\sim+\infty)=0\ \ \Leftrightarrow\ \ \left\{\begin{array}{rcl}\displaystyle\alpha_0(r_1\sim+\infty)&=&\displaystyle-1+M_0\,M_N\left(\sum_{n=1}^NM_n^2\right)^{-1}\ , \\[5mm]\displaystyle\alpha_n(r_1\sim+\infty)&=&\displaystyle-1+M_n\,M_N\left(\sum_{n=1}^NM_n^2\right)^{-1}\ .\end{array}\right.
\label{alpha_limits}
\end{equation} 
These limits constitute a consistency check for the map~(\ref{map}) and the definition of the reference profile $\alpha_0$, in particular
allow to set the asymptotic limit of such profile for large radii.

We note that when considering a detailed model containing the known massive bodies within a given gravitational system the ansatz~(\ref{metric_ELA_N}) describing each individual body contribution to the background is applicable requiring a full numerical simulation with $N$ functional parameters. However for a simplified model considering a planar average mass surface density, the ansatz~(\ref{metric_ELA_eff}) describing one single effective functional parameter on the galaxy plane reduces significantly the complexity of the model allowing for a simpler parameterization of gravitational corrections. Hence in the remaining of this work we will employ the effective metric ansatz~(\ref{metric_ELA_eff}) to develop a simplified disk lattice model for galaxies similar to cosmological lattice models~\cite{gal_lat}.
For a given galaxy, once a discretized $2D$ profile for the functional parameter $\alpha_{\mathrm{eff}}$ is obtained by fitting the rotation curve on the plane of the galaxy, the discretized profile for the functional parameter of each body on the galaxy model, $\alpha_n$~(\ref{metric_ELA_N}), can be parameterized by the scaling law for $\alpha_M$~(\ref{scaling_M}) such that solving the system of equations corresponding to map~(\ref{map}) we obtain the discretized profile for $\alpha_0$ on the galaxy plane corresponding to the model considered. From this discretized profile it is straight forward to explicitly compute the discretized profile for each of the $\alpha_n$, also on the galaxy plane, from the scaling law~(\ref{scaling_M}). Further assuming either spherical symmetry (or a parameterization of the anisotropy) of the parameter $\alpha_n$ with respect to the center of mass of each body in the model, the value for the functional parameter $\alpha_{\mathrm{eff}}$ outside of the galaxy plane is computed by employing the map~(\ref{map}), hence allowing for a $3D$ analysis of derived gravitational quantities in the neighborhood of the galaxy such as mass-energy density, anisotropic pressures and equation of state corrections due to the ELA metric background. 

When computing the equations of motion for either the metric~(\ref{metric_ELA_eff}) or~(\ref{metric_ELA_N}), there is one more problem to address, although for the SC metric we have considered the series expansion to first order in the gravitational field for the metric components $g_{SC.00}=1-U_N$ and $g_{SC.rr}=1$, we have not considered a series expansion for the shift function $N^r_{ELA}$. Although for small values of the metric functional parameter $|\alpha|\sim 10$ an expansion of the factor $(1-U)^\alpha$ to the same order of the metric component $g_{00}$ maintains the accuracy of the ELA metric approximation, for higher values of the functional parameter a higher order expansion of the shift function on the gravitational field $U$ is required to attain the same accuracy. We recall that a similar problem is verified in the PN formalism as the component $g_{00}$ requires an higher order expansion than the remaining metric components~\cite{PN}. Although an higher order expansion of the ELA metric shift function can be considered, there is no clear technical advantage as it will difficult the derivation of the equations of geodesic motion and other derived quantities. Instead, for technical simplification we will carry the following computations to first order on the gravitational field with respect to the local metric background components $g_{SC.00}=1-U_N$ and $g_{SC.rr}=1$ while considering the exact expression for the shift function $N^r_{ELA}$. Also this construction can be justified by noting that the shift function is here a functional parameter, hence a generic function for which the specific dependence on the gravitational field is unknown such that the specific order of its series expansion in the gravitational field cannot be exactly determined. In addition we will consider a series expansion for the metric and remaining derived quantities of second order on the Hubble rate $H$, hence of second order on the shift function $N^r_{ELA}$.

Given this modeling setup we proceed to compute the metric connections and obtain the dominant contributions to the radial acceleration of a test mass on the many body background described by the ansatz~(\ref{metric_ELA_eff}). Considering that the many body system lies approximately on a plane of constant $\theta$ and for non-relativistic velocities $\dot{r}_1\ll c$, $\dot{\varphi}_1\ll c$ and $\dot{\theta}_1\ll c$, the radial acceleration approximated to first order on the gravitational field $U_N$ and second order on the Hubble rate $H$ is
\begin{equation}
\begin{array}{rcl}
\ddot{r}_1&\approx& -c^2\,\Gamma^{r}_{\ tt}-\Gamma^{r}_{\varphi\varphi}\,\dot{\varphi}^2\approx F_N+F_{\varphi}+F_{H^2}+O(U_N^2,H^4)\ ,\\[5mm]
F_N&=&\displaystyle\frac{c^2}{2}U'_N\ ,\ F_{\varphi}=r_1\,\dot{\varphi}^2\,\sin\theta\\[5mm]
F_{H^2}&=&\displaystyle -\frac{c^2}{2}\,\left(N^r_{ELA.\mathrm{eff}}\right)^2\,U'_N\\[5mm]
&&\displaystyle+c^2N^r_{ELA.\mathrm{eff}}\,N'^r_{ELA.\mathrm{eff}}+c\,\dot{N}^r_{ELA.\mathrm{eff}}\\[5mm]
&=&\displaystyle\frac{H^2r_1}{2}\,\left(1-U_N\right)^{\alpha_{\mathrm{eff}}}\Bigl((1-U_N)\left(2+r_1\log(1-U_N)\alpha'_{\mathrm{eff}}\right)\\[4mm]
&&\displaystyle-2(1+q)(1-U_N)^{\frac{1}{2}-\frac{\alpha_{\mathrm{eff}}}{2}}-r_1(2+\alpha_{\mathrm{eff}}-U_N)\,U'_N\Bigr)\ ,
\end{array}
\label{F_1}
\end{equation}
where dotted quantities represent differentiation with respect to the time coordinate $t$ and primed quantities differentiation with respect to the radial coordinate $r_1$. The acceleration component $F_N$ is the standard gravitational Newton acceleration, $F_\varphi$ is the standard centripetal acceleration and $F_{H^2}$ is the lower order correction in the Hubble rate $H$ due to the expanding locally anisotropic background as described by the ELA metric~(\ref{metric_eff}).

As long as the parameter $\alpha_{\mathrm{eff}}$ is finite, for small values of $r_1$, $F_{H^2}\approx 0$ is negligible, while for large enough values of $r_1$ the acceleration component $F_{H^2}$ is positive being
approximated by the usual RW acceleration $\lim_{r_1\to\infty}F_{H^2}\approx F_{RW}=-qH^2\,r_1$, hence coinciding with the cosmological
acceleration outwards the observer. In between these two asymptotic limits, depending on the value of the functional parameter $\alpha_{\mathrm{eff}}$, the component $F_{H^2}$ may either be positive or negative. We note that, by numerical inspection of $F_{H^2}$, for large negative values of the parameter $\alpha_{\mathrm{eff}}$ the correction to the Newton gravitational acceleration is negative, hence towards the central mass~\cite{Pioneer}. In particular the metric functional parameter allows for a covariant parameterization of the excess acceleration towards the core of galaxies, hence increasing the orbital speed of massive bodies on the galaxy and allowing to describe the experimentally observed flattening of galaxy rotation curves~\cite{v_gal1,v_gal2}. In the next section we explicitly derive a discrete disk galaxy model approximately matching the physical parameters of the galaxy UGC2885.

As for the mass-energy density for the many body ELA metric~(\ref{metric_ELA_eff}) it is 
\begin{equation}
\begin{array}{rcl}
\rho_\alpha&=&\displaystyle\frac{c^2}{8\pi\,G\,r_1^2}(1-U_N)\Biggl(N^r_{ELA.\mathrm{eff}}\left(N^r_{ELA.\mathrm{eff}}+2\,r_1\,\partial_rN^r_{ELA.\mathrm{eff}}\right)\\[5mm]
&&\displaystyle-\frac{1}{4}\left(\left(\partial_\theta N^r_{ELA.\mathrm{eff}}\right)^2+\frac{1}{\sin^2\theta}\left(\partial_\varphi N^r_{ELA.\mathrm{eff}}\right)^2\right)\Biggr)\\[6mm]
&=&\displaystyle \frac{3\,H^2}{8\pi\,G}\,(1+U_N)(1-U_N)^{\alpha_{\mathrm{eff}}-1}\Biggl((1-U_N)^2\times\\[5mm]
&&\displaystyle\times\left(1+\frac{r_1}{3}\log(1-U_N)\partial_r\alpha_{\mathrm{eff}}\right)-\frac{1-\alpha_{\mathrm{eff}}}{48}\Bigl((1-\alpha_{\mathrm{eff}})(\partial_\theta U_N)^2\\[5mm]
&&\displaystyle+\frac{1-\alpha_{\mathrm{eff}}}{\sin^2\theta}(\partial_\varphi U_N)^2+16 r_1(1-U_N)\partial_r U_N \Bigr)\Biggr)
\end{array}
\label{rho_alpha}
\end{equation}
In the limit of large radii, as long as $\alpha_{\mathrm{eff}}$ is finite, this expression coincides with the RW mass-energy density $\rho_{RW}=3H^2/(8\pi G)$.

\section{A Lattice model for galaxies}

In this section we will apply the many body ELA metric~(\ref{metric_ELA_eff}) to develop a simplified lattice disk model for galaxies
showing that the flattening of galaxy rotation curves can be fully parameterized by this ansatz.
To model the galaxy baryonic matter we are considering a discrete polar lattice on the galaxy plane with point-like masses at each lattice
vertex. Although it is possible to implement an analytical model on the plane of the galaxy, to compute the values of the effective
functional parameter $\alpha_{\mathrm{eff}}$~(\ref{map}) outside the galaxy plane employing the method described in the previous section it is
required to consider a gravitational system of point like masses as it employs a reference profile $\alpha_0$ for
each massive body~(\ref{scaling_M}). We also note that an explicit discretization of the mass-energy density simplifies the computational simulation of
the galaxy model as we will develop in the remaining of this section. To explicitly compute the values of the point-like masses on galactic plane we consider the same model employed in~\cite{mass_1} consisting of an infinitesimal $2D$ thin exponential disk~\cite{disk} describing the intergalactic gas and a $3D$ thin exponential disk~\cite{disk_3D} describing stellar matter across the galaxy plane plus a bulge matter distribution near the center of the galaxy describing the galaxy core. The galaxy disk is considered to be finite having an explicit cut-off describing the galaxy edge~\cite{disk_cutoff}.

For exemplification purposes we are modeling the galaxy UGC2885. Experimental photometric data and red-shift measurements
for this galaxy are available in the works~\cite{v_gal1,v_gal2,v_gal3,v_gal3a,v_gal4}. We note that within the analysis
carried in these references there are discrepancies in the derived physical parameters of the galaxy. These discrepancies are mainly due to the distinct estimates for the Heliocentric distance $D$, which is derived from the systemic velocity $V_{sys}$ which, in turn, is computed from the average of the measured red-shift. While in~\cite{v_gal1,v_gal2} it is considered the uncorrected systemic velocity $V_{sys}=5794\,km\,s^{-1}$ corresponding to a Heliocentric distance of $D=118\,Mpc$, in~\cite{v_gal3a,v_gal3} the systemic velocity is converted to the motion relative to the local group and the cosmic microwave background $\bar{V}_{sys}=5683\,km\,s^{-1}$ corresponding to the Heliocentric distance of $D=76\,Mpc$. In addition, given a estimate for $\bar{V}_{sys}$, the estimate for $D$ also depends on the today's Hubble rate value $H_0$ and deceleration parameter $q_0$ which has been updated~\cite{wmap} with respect to the values considered in these references. The mass modeling of the galaxy is sensitive to several of the
physical parameters, hence we re-derive these parameters considering the converted systemic velocity $\bar{V}_{sys}=5683\,km\,s^{-1}$~\cite{v_gal3,v_gal3a}
and today's value for the Hubble rate $H_0=2.28\times 10^{-18}\, s^{-1}$ and deceleration parameter $q_0=-0.582$. The derived physical parameters
for the galaxy UGC2885 are listed in table~\ref{table.gal_prop}.
\begin{table}[ht]
\begin{center}
\begin{tabular}{llll}
parameter& & value &ref \\\hline\hline\\
$M_{\mathrm{tot}}$&total mass&$1.3\times 10^{12}\,M_\odot$&\cite{v_gal3}\\
$M_{\mathrm{tot}}/L_B$&mass to light ratio&$4.7\,M_\odot/L_\odot$&\cite{v_gal1}\\
$M_{H_I}/M_{\mathrm{tot}}$&fractional $H_I$ mass&$\approx 2.3\%$&\cite{v_gal2}\\
$M_{DB}/M_{\mathrm{tot}}$&disk fractional stellar mass&$\approx 21\%$&\cite{v_gal2,v_gal3}\\
$V_{rot.lim}$&rotational velocity at $R_{max}$&$298\,(km\,s^{-1})$&\cite{v_gal1,v_gal2}\\
$V_{sys}$&measured systemic velocity&$5794\,(km\,s^{-1})$&\cite{v_gal3}\\
$\bar{V}_{sys}$&converted systemic velocity&$5683\,(km\,s^{-1})$&\cite{v_gal3}\\
$R_{max}$&galaxy semi-major axis&$83.89\,(kpc)$&\cite{v_gal1,v_gal2,v_gal3}\\
$z$&cosmological red-shift&$0.01896$&\cite{wmap}\\
$D$&Heliocentric distance&$81.14\,Mpc$&\cite{wmap}\\
$H$&Hubble rate&$2.26\times 10^{-18}\,s^{-1}$&\cite{wmap}\\
$q$&deceleration parameter&$-0.584$&\cite{wmap}\\
\\\hline
\end{tabular}
\end{center}
\caption{\small Physical parameters of the galaxy UGC2885. The systemic velocity $\bar{V}_{sys}$ is converted to the heliocentric velocity
with respect to the local group and the cosmic microwave background~\cite{v_gal3} and the remaining quantities $R_{max}$, $z$, $D$, $H$ and $q$, are computed for the most recent estimates for $H_0$ and $q_0$~\cite{wmap}.\label{table.gal_prop}}
\end{table}

Following the modeling setup discussed in~\cite{mass_1}, to explicitly define the gas and stellar matter surface densities on the galaxy plane we are considering the neutral Hydrogen $H_I$ $21\,cm$-line photometric measurements analyzed in~\cite{v_gal2} and the K-band photometric measurement analyzed in~\cite{v_gal3a}. The gas matter surface density is modeled by a thin $2D$ exponential disk matching the $H_I$ matter surface density multiplied by the corrective factor of $1.4$~\cite{mass_1} and the stellar matter surface density is modeled by integrating along the orthogonal spatial direction to the galaxy plane the $3D$ exponential disk matching the K-band luminosity in~\cite{mass_1}. As for the galaxy bulge we directly fit the excess mass required to the rotation curve up to $28\,kpc$, hence describing the rising of this curve near
the galactic core. In addition, due to the distinct estimates for $D$ in~\cite{v_gal2} and~\cite{v_gal3a}, it is required to scale the surface densities computed in these references accordingly. Hence we obtain
\begin{equation}
\begin{array}{rcl}
\mu&=&\displaystyle\mu_D+\mu_B+\sum_{i=0}^{N_{28}}\mu_{S[i]}\ ,\\[5mm]
\mu_D&=&\displaystyle 0.00791432\,e^{-\left(\frac{r_1}{37.5662\,kpc}\right)^2}\delta_{\mathrm{cut-off}}(r_1)\hfill(kg\,m^{-2}),\\[5mm]
\mu_B&=&\displaystyle 0.329868\,e^{+\frac{r_1}{11.6367\,kpc}}\delta_{\mathrm{cut-off}}(r_1)\hfill(kg\,m^{-2}),\\[5mm]
\delta_{\mathrm{cut-off}}(r_1)&=&\left\{\begin{array}{lcl}1&,&r_1\leq R_{max}\\ \displaystyle 1-\frac{r_1-R_{max}}{\delta}&,&R_{max}<r_1< R_{max}+\delta\\0&,&r_1\geq R_{max}\end{array}\right.
\end{array}
\label{mu}
\end{equation}
where, for a given radial coordinate discretization $r_{1[i]}$, $\mu_{S[i]}$ are constant surface density disks with edge
at $r_1=(r_{1[i]}+r_{1[i+1]})/2$, $N_{28}$ corresponds to the integer labeling
of the discretized radial coordinate nearest to $28\,kpc$ and the function $\delta_{\mathrm{cut-off}}$ describes a smooth cut-off for the galaxy edge near $R_{max}$.

We now proceed to explicitly define a lattice model for the galaxy UGC2885 similar to models considered for many body cosmological simulations~\cite{gal_lat}. Aligning the $z$ axis orthogonally to the galaxy plane such that the galaxy disk lays in the plane of constant $\theta=\pi/2$ and
considering a regular polar lattice discretization over the plane of the galaxy by $i_{max}$ points along the radial coordinate
$r_1$ and $k_{max}$ points along the angular coordinate $\varphi$ we obtain the discrete polar coordinates for the lattice points labeled by $[i,k]$
\begin{equation}
\begin{array}{rcl}
r_{1[i]}&=&\displaystyle i\times \Delta_{r_1}\ ,\ i=0,1,\ldots, i_{max}\ ,\\[5mm]
\varphi_{[k]}&=&\left\{\begin{array}{lcl}2\pi&,&i=0\\ \displaystyle i\neq 0\ ,\ k\times \Delta_k&,&k=\displaystyle 0,1,\ldots, k_{max}-1\ \ ,\ \  \Delta_k=\frac{2\pi}{k_{max}}\ .\end{array}\right.\\[5mm] 
\end{array}
\end{equation}
The lattice faces are centered at each lattice point $[i,k]$, hence being defined as bounded $2D$ surfaces, and at each lattice point $[i,k]$ we consider a point-like mass $M_{[i]}$ of value matching the integrated surface mass-energy density over the lattice $\mathrm{face}_{[i,k]}$
\begin{equation}
\begin{array}{rcl}
\mathrm{face}_{[i,j]}&=&\displaystyle\left[r_{1[i]}-\frac{\Delta_{r_1}}{2},r_{1[i]}+\frac{\Delta_{r_1}}{2}\right[\times\left[\varphi_{1[k]}-\frac{\Delta_k}{2},\varphi_{1[k]}+\frac{\Delta_k}{2}\right[\ ,\\[5mm]
M_{[0]}&=&\displaystyle 2\pi\int_{0}^{\frac{\Delta r_1}{2}} dr_1\,r_1\,\mu(r_1)\ ,\\[5mm]
M_{[i]}&=&\displaystyle\Delta_k\int_{r_{1[i]}-\frac{\Delta_{r_1}}{2}}^{r_{1[i]}+\frac{\Delta_{r_1}}{2}} dr_1\,r_1\,\mu(r_1)\ ,
\end{array}
\label{M_i}
\end{equation}
where $\mu$ is the modeled surface mass density~(\ref{mu}). Hence we are considering $N=1+i_{max}\times k_{max}$ point-like massive bodies over the planar polar lattice.

\begin{figure}[!htbp]
\begin{center}
\includegraphics[width=120mm]{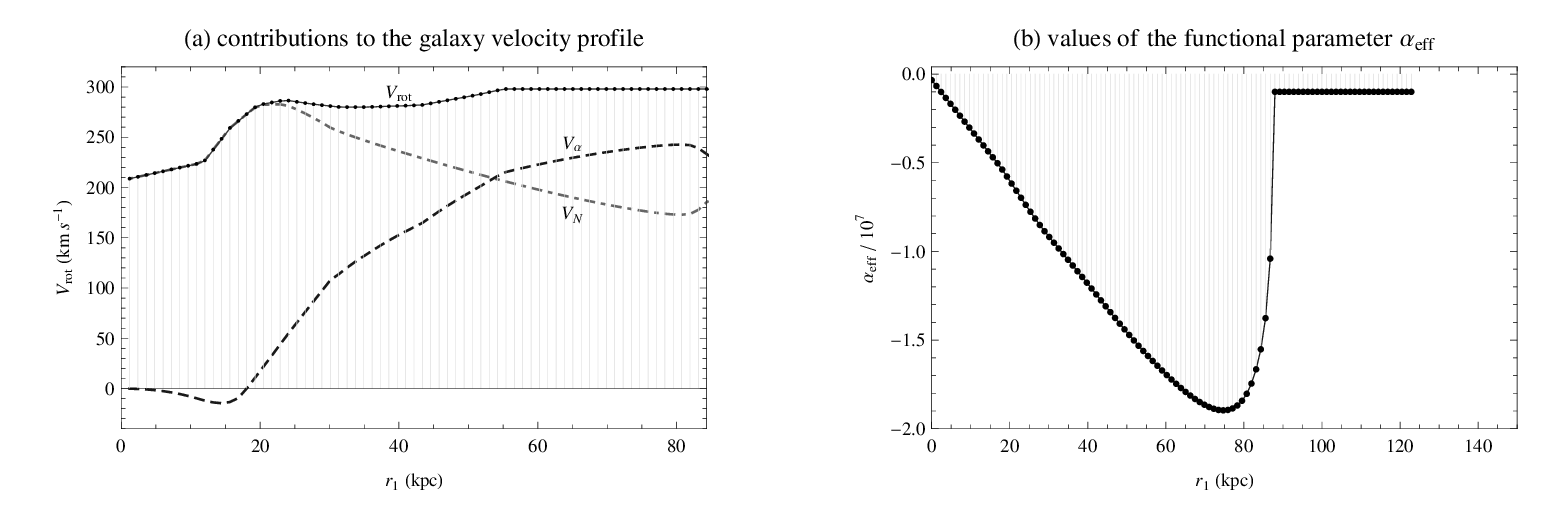}
\caption{\small {\bf (a)} Velocity profiles for UGC2885. The velocity contribution due to the Newton acceleration ($V_{N}$) is represented by a dashed-dot line,
the velocity contribution due to ELA metric correction to the gravitational acceleration ($V_\alpha$) is represented by a dashed line
and the total orbital velocity ($V_{\mathrm{orb}}$) by a continuous line with the lattice points represented by dots; {\bf (b)} profile for the functional parameter $\alpha_{\mathrm{eff}}$ on the galactic plane, the lattice points are represented by dots.\label{fig.V}}
\end{center}
\end{figure}

Assuming that all the bodies are approximately at a stable circular orbit the galaxy model exhibits explicit planar circular symmetry on the plane of galaxy such that both the orbital velocities and the gravitational potentials for the bodies at each lattice point $[i,k]$ are independent of the angular coordinate index $k$. Hence, noting that the time derivative of the angular coordinate $\varphi$ for each value of the radial coordinate $r_{1[i]}$ is $\dot{\varphi}_{\mathrm{orb}[i]}=V_{\mathrm{orb}[i]}/r_{1[i]}$, the radial equations of motion~(\ref{F_1}) on the galaxy plane are
\begin{equation}
\ddot{r}_{1[i]}=0\ \ \Leftrightarrow\ \ V_{\mathrm{orb}[i]}=\sqrt{-r_{1[i]}(F_{N[i]}+F_{H^2[i]})}\ ,
\label{V_orb}
\end{equation}
and the the potential $U$ and its derivative $U'$ have the following explicit expressions at each lattice point
\begin{equation}
\begin{array}{rcl}
U_{[i]}&=&\displaystyle\sum_{[j,k]\neq [i,0]}\frac{2GM_{[i]}}{c^2\Delta r_{1[i,0][j,k]}}\ ,\\[6mm]
U'_{[i]}&=&\displaystyle-\sum_{[j,k]\neq [i,0]}\frac{2GM_{[i]}(r_{1[i]}-r_{1[j]}\cos(\varphi_{[k]}))}{c^2(\Delta r_{1[i,0][j,k]})^3}\ ,\\[6mm]
\Delta r_{1[i,0][j,k]}&=&\sqrt{r_{1[i]}^2-2r_{1[i]}r_{1[j]}\cos(\varphi_{[k]})+r_{1[j]}^2}\ .
\end{array}
\label{U_dU_i}
\end{equation}
In these expressions we are considering as the test masses the bodies at the lattice points $[i,0]$. Due to the circular symmetry of the
planar model any other value of $k$ for the test masses may generally be considered. As for the functional parameter $\alpha_{\mathrm{eff}}$ on the galactic plane we consider it to be a radial symmetric function such that its values is also independent of the angular lattice index $k$.
In the following we will fit this parameter to the experimentally observed galaxy rotation curves, hence it is further required to consider a numerical derivative to evaluate the derivative of this parameter with respect to the radial coordinate $\alpha'_{\mathrm{eff}}$. We consider
a first order approximation to this derivative. To reduce the numerical uncertainty on this approximation we further consider the parameterization of the functional parameter by a second order expansion on the gravitational field, $\alpha_{\mathrm{eff}}=3+\alpha(1-U)^2$. This parameterization does not necessarily has any physical meaning, it is employed here as a parameterization that effectively reduces the relative magnitude of the contribution of the first order discrete derivative, hence allowing to numerically fit the functional parameter of the model to the existing velocity profiles without requiring to
consider an higher order approximation to the derivative. Therefore, at each lattice point $[i,k]$ on the plane of the galaxy (again independently of the index $k$) we obtain the following expressions for the functional parameter and its derivative with respect to the radial coordinate
\begin{equation}
\begin{array}{rcl}
\alpha_{\mathrm{eff}[i]}&=&3\displaystyle+\alpha_{[i]}(1-U_{[i]})^2\ ,\\[5mm]
\alpha'_{\mathrm{eff}[i]}&=&\displaystyle-2U'_{[i]}(1-U_{[i]})+(1-U_{[i]})^2\frac{\alpha_{[i+1]}-\alpha_{[i]}}{r_{1[i+1]}-r_{1[i]}}\ ,
\end{array}
\label{alpha_eff_i}
\end{equation}
where we have considered forward finite differences when writing the discrete derivative of the functional coefficient $\alpha_{[i]}$.
This construction allows to describe the unknown functional parameter that parameterizes the gravitational interactions corrections
by its values across the discretized lattice points. As for the gravitational acceleration contributions to the orbital
velocity~(\ref{V_orb}) are
\begin{equation}
\begin{array}{rcl}
F_{N[i]}&=&\displaystyle-\frac{c^2}{2}U'_{[i]}\ ,\\[5mm]
F_{H^2[i]}&=&\displaystyle \frac{H^2 r_{1[i]}}{2} \left(1 - U_{[i]}\right)^{\alpha_{[i]}} \Bigl(2 \left(1 - U_{[i]}\right)\\[5mm]
&&\displaystyle-2(1+q)\left(1-U[i]\right)^{\frac{1}{2} - \frac{\alpha_{[i]}}{2}}\\[5mm]
&&\displaystyle -r_{1[i]} \left(2 + \alpha_{[i]} - U_{[i]}\right) U'_{[i]}\Bigr)+\Delta F_{H^2[i]}\\[6mm]
\Delta F_{H^2[i]}&=&\displaystyle\frac{(H\,r_{1.[i]})^2}{2}(1-U_{[i]})^{\alpha_{\mathrm{eff}[i]}+1}\log(1-U_{[i]})\alpha'_{\mathrm{eff}[i]}\ .
\end{array}
\label{dF}
\end{equation} 
$V_{\mathrm{orb}[i]}$~(\ref{V_orb}) can now be fitted to the experimental measured rotational velocities. At each lattice point the values for the
experimentally measured velocity $V_{\mathrm{exp}[i]}$
are linearly interpolated for each value of the radial coordinate $r_{1[i]}$ to the experimental profile listed in table~3 of reference~\cite{v_gal2}.
To numerically evaluate the values of the functional parameter $\alpha_{\mathrm{eff}}$ that fit this data we are
considering an edge smoothing of $\delta=5\,kpc$ and a discretization of the radial coordinate on the galaxy plane up to $R_{max}+\delta=88.89\,kpc$ with the following
lattice properties
\begin{equation}
\begin{array}{rcl}
i_{max}&=&74\ ,\\[5mm]
k_{max}&=&16\ ,\\[5mm]
\Delta_{r_1}&=&1.20464\,(kpc)\ ,\\[5mm]
\Delta_{k}&=&\displaystyle\frac{\pi}{8}\ ,\\[5mm]
N&=&1185\ .
\end{array}
\end{equation}
For values of $\alpha_{\mathrm{eff}[i]}\sim -10^7$ on the galaxy disk the flattening of the galaxy rotation curve is fully accounted for. The velocity contributions due to the Newton gravitational acceleration $V_{N[i]}=\sqrt{-r_{1[i]}F_{N[i]}}$ and due to the background described by the ElA metric $V_{\alpha[i]}=\sqrt{-r_{1[i]}(F_{H^2[i]}+\Delta F_{H^2[i]})}$ to the total orbital velocity $V_{\mathrm{orb}[i]}=\sqrt{V_{N[i]}^2+V_{\alpha[i]}^2}$ are plotted in figure~\ref{fig.V}. The values for these velocities as well as the values of the point-like baryonic masses $M_{i}$ and the coefficients
$\alpha_{[i]}$ are listed in table~\ref{table.V} in the appendix.
We note that reducing the spacing of the lattice, hence increasing the number of the bodies in 
the lattice model and reducing the mass for each of these massive bodies, does not change significantly
the value of the effective functional parameter $\alpha_{\mathrm{eff}}$. This is mainly due to the shift
function depending on the total gravitational potential $U_N$ which is approximately maintained constant
for each value of the radial coordinate independently of how many discretization points have been considered.

The total baryonic mass for this model is
\begin{equation}
M_b=2.30\times 10^{11}\,M_\odot\ .
\label{M_b}
\end{equation}
However there are contributions to the total galaxy matter due to
mass-energy density contribution due to the expanding locally anisotropic background corrections with respect to Schwarzschild backgrounds~\cite{Pioneer}. In particular it is relevant to remark that to ensure that causality is preserved the total mass-energy density must be
strictly positive both on the galaxy disk as well as beyond the galaxy edge.
For the particular fit discussed here we have considered that beyond the galaxy disk edge the value of
the parameter $\alpha_{\mathrm{eff}}$ is approximately a constant $\sim -10^{-6}$ ensuring that the mass-energy density of the background is strictly positive. Next we analyze and discuss in detail the mass-energy densities for the galaxy model.

\section{Mass-energy density analysis}

\begin{figure}[!htbp]
\begin{center}
\includegraphics[width=120mm]{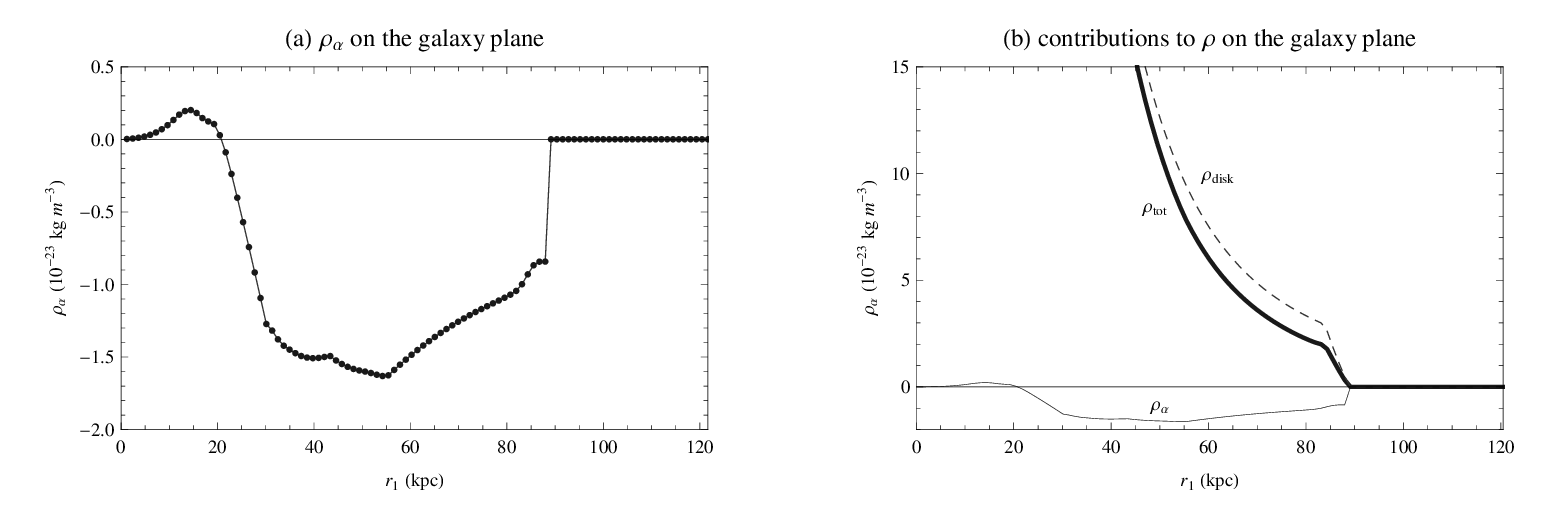}
\caption{\small Mass-Energy density profiles for the galaxy model: {\bf (a)} mass-energy density $\rho_{\alpha}$~(\ref{rho_alpha}) on the galactic plane, the lattice point are represented by dots; {\bf (b)} contributions $\rho_{\alpha}$ (thin line) and $\rho_{\mu}$ (dashed line) to the total mass-energy density $\rho_{tot}$ (thick line) on the galaxy plane. The total mass-energy density is strictly positive.\label{fig.rho}}
\end{center}
\end{figure}

To evaluate the mass-energy density $\rho_\alpha$ due to the locally anisotropic background within the galaxy plane it is enough to evaluate the expression~(\ref{rho_alpha}). We have already fitted the functional parameter $\alpha_{\mathrm{eff}}$ to the rotation curve on the plane of the galaxy as pictured in figure~\ref{fig.V}, hence it is straight forward to evaluate $\rho_{\alpha[i,k]}$ at each lattice point $[i,k]$ by excluding the point-like
mass at this lattice point and consider the contribution of the remaining $N-1$
point-like masses. Noting that the derivatives of the gravitational potential $\partial_\theta U$ and $\partial_\varphi U$ are null and that $\rho_{\alpha}$ is circular symmetric on the galaxy plane, the discrete quantities $U_{[i]}$, $U'_{[i]}$~(\ref{U_dU_i}), $\alpha_{\mathrm{eff}}$ and $\alpha'_{\mathrm{eff}}$~(\ref{alpha_eff_i}) already computed are enough to actually compute the profile for the mass-energy density
$\rho_{\alpha[i,0]}$ on the galaxy plane. Such profile is plotted in figure~\ref{fig.rho}.

We can readily verify that there is a negative mass-energy density contribution due
to the expanding locally anisotropic background for radial distances greater than $21\,kpc$ from the center of the galaxy. Strictly negative mass-energy densities violate causality, therefore are commonly not considered
as a physical reality. For the specific model discussed here the mass-energy density is strictly positive on the galaxy plane up to the distance of $89\,kpc$ from the galaxy center such that causality is preserved. To show it explicitly let us note that the surface mass-energy density
contributions $\mu_B$ and $\mu_D$~(\ref{mu}) are effective 2D 
quantities describing the 3D mass-energy densities $\rho_B$
and $\rho_D$, respectively. Specifically $\rho_B$ is fitted to a exponential 3D disk in~\cite{v_gal3a} and for $\rho_D$ we assume
that it is approximately described by a Gaussian disk of thickness
$\Delta_{r_1}/2$
\begin{equation}
\begin{array}{rcl}
\rho_B&=&4.60183\times 10^{-21}\,e^{-\frac{r_1}{11.6367\,kpc}-\frac{z}{1.16367\,kpc}}\ ,\\
\rho_D&=&2.40691\times 10^{-22}\,e^{-\frac{r_1}{37.5662\,kpc}-\left(\frac{z}{0.602322\,kpc}\right)^2}\ .
\end{array}
\end{equation}
Hence within the galaxy disk the sum of these two contributions plus the contribution $\rho_\alpha$~(\ref{rho_alpha}) is strictly positive. However, beyond the galaxy disk edge $r_1\geq 89\,kpc$, the baryonic mass-energy density is approximately null such that the only contribution to the total mass-energy density is due to $\rho_\alpha$. For the galaxy model presented here $\rho_\alpha>0$ for $r_1>89\,kpc$ when $\alpha_{\mathrm{eff}}\ge-10^6$. Also we recall that $\rho_\alpha$ does contribute to the total mass of the galaxy in addition to the baryonic mass such that to ensure a finite contribution mass for the model we further consider a upper radial cut-off $R_\alpha$ above which the functional parameter is null $\alpha_{\mathrm{eff}}=0$ such that explicit spatial isotropy is recovered and the mass-energy density above this cut-off exactly matches the one for the expanding Universe $\rho_{RW}$~\cite{McV}. Hence both to preserve causality beyond the galaxy edge and to ensure a total finite galaxy mass we consider that the functional parameter $\alpha_{\mathrm{eff}}$ is approximately a constant in-between the radial distance of $89\,kpc$ and $R_\alpha$
\begin{equation}
\alpha_{\mathrm{eff}} =\left\{\begin{array}{lcl} -10^6&,&r_1\in\left]88.89\,kpc,R_\alpha\right]\ ,\\[5mm]0&,&r_1\in\left]R_\alpha,+\infty\right[\ .\end{array}\right.
\label{alpha_big}
\end{equation}
This ansatz is based only on the assumption that causality must be maintained as well as the total galaxy mass must be finite. A more realistic setup
and estimate for $\alpha_{\mathrm{eff}}$ would require the analysis of direct observational data outside the galaxy disk, for instance the gravitational lens effect on the background radiation due to the deformation of the gravitational background induced by the galaxy. We leave such analysis to another work, here we will proceed our analysis taking the ansatz~(\ref{alpha_big}) as an approximately constant lower cut-off for
the functional parameter $\alpha_{\mathrm{eff}}$.

Given the model setup just described, the total mass-energy density is strictly positive on the galaxy plane both within the galaxy disk and beyond the galaxy edge. The contributions to the total mass-energy density on the galaxy plane are plotted in figure~\ref{fig.rho}.

In addition to the analysis of the mass-energy density contributions and the ansatz for the value of $\alpha_{\mathrm{eff}}$ beyond the galaxy edge one is led to the question whether our model can further be extended in a 3D neighborhood of the galaxy, in particular whether it is possible to estimate the value of the functional parameter and the mass-energy density along the orthogonal direction to the plan of the galaxy. To implement such analysis let us consider the map between metric~(\ref{metric_ELA_eff}) and metric~(\ref{metric_ELA_N}) and assume the existence of an unique reference profile $\alpha_{0}$ corresponding to a reference mass of $M_0$ such that each for the functional parameters $\alpha_n$ are defined with respect to $\alpha_0$ accordingly to the scaling law~(\ref{scaling_M}). If the 3D profile for $\alpha_{0}$ is known also the 3D profiles for the several $\alpha_n$'s can be computed from the scaling law~(\ref{scaling_M}) and the value for the parameter $\alpha_{\mathrm{eff}}$ along the orthogonal direction to the galaxy plan can be estimated from the map~(\ref{map}) by considering some given symmetry for the several
functional parameters $\alpha_n$.

Noting that the galaxy model is explicitly circular symmetric, a discrete planar profile $\alpha_{0[i]}$ on the galaxy plane for the functional parameter $\alpha_0$ can be computed by solving the linear system
\begin{equation}
\alpha_{\mathrm{eff}[i]}=-1+\frac{\displaystyle\sum_{[j,k]\neq[i,0]}\frac{M_{[j]}}{M_0}\left(\tilde{\alpha}_{0[i,0][j,k]}+1\right)\log\left(1-U_{n[i,0][j,k]}\right)}{\log\left(1-U_{[i]}\right)}\ ,
\label{eff_n_sys}
\end{equation}
where the potentials $U_{n[i,0][j,k]}$ correspond to the gravitational potential of the point-like massive body located at the lattice point $[j,k]$ evaluated at the lattice point $[i,0]$ and $\tilde{\alpha}_{0[i,0][j,k]}$ is the linear interpolated profile for $\alpha_{0[i]}$ evaluated
at a distance $\Delta r_{1[i,0][j,k]}$ from $\alpha_{0[0]}$ 
\begin{equation}
\begin{array}{rcl}
U_{n[i,0][j,k]}&=&\displaystyle\frac{2GM_{[j]}}{c^2\,\Delta r_{1[i,0][j,k]}}\ ,\\[7mm]
\tilde{\alpha}_{0[i,0][j,k]}&=&\left\{\begin{array}{l}
\alpha_{0[\tilde{\jmath}]}+\frac{(\Delta r_{1[i,0][j,k]}-r_{1[\tilde{\jmath}]})(\alpha_{0[\tilde{j}]}-\alpha_{0[\tilde{\jmath}+1]})}{\Delta_{r_1}}\ ,\\ \hfill\mathrm{for}\ \Delta r_{1[i,0][j,k]}\geq r_{1[\tilde{\jmath}]}\\[5mm] 
\alpha_{0[\tilde{\jmath}-1]}+\frac{(\Delta r_{1[i,0][j,k]}-r_{1[\tilde{\jmath}-1]})(\alpha_{0[\tilde{j}-1]}-\alpha_{0[\tilde{\jmath}]})}{\Delta_{r_1}}\ ,\\
\hfill\mathrm{for}\ \Delta r_{1[i,0][j,k]}<r_{1[\tilde{\jmath}]}
\end{array}\right.\\[15mm]
r_{1[\tilde{\jmath}]}&=&\mathrm{nearest}(\Delta r_{1[i,0][j,k]})\ .
\end{array}
\label{alpha_0_int}
\end{equation}
The potentials $U_{[i]}$ and the distances $\Delta r_{1[i,0][j,k]}$ are defined in~(\ref{U_dU_i}) and $\tilde{\jmath}$ is
the index of the discrete radial coordinate nearest to $\Delta r_{1[i,0][j,k]}$.

\begin{figure}[!htbp]
\begin{center}
\includegraphics[width=60mm]{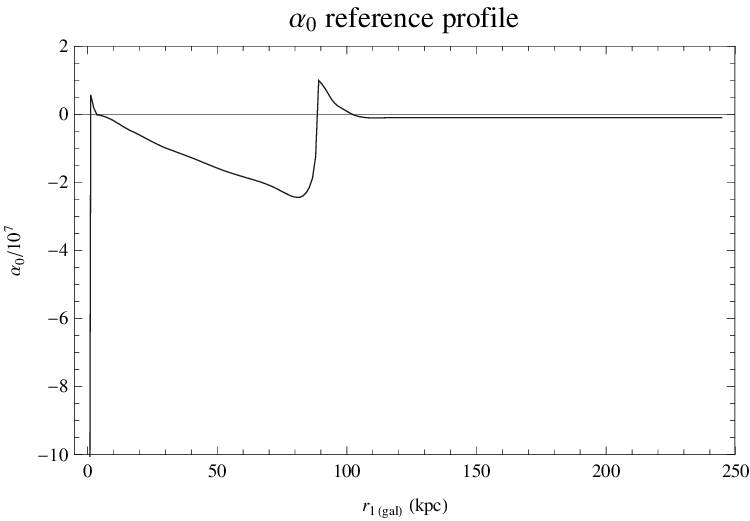}
\caption{\small Discrete circular symmetric reference profile $\alpha_0$ on the galaxy plane. \label{fig.alpha_0}}
\end{center}
\end{figure}

To ensure that the system of equations~(\ref{eff_n_sys}) is not degenerate it is required that both the profile $\alpha_{\mathrm{eff}[i]}$ and
$\alpha_{0[i]}$ have the same length such that the indice $i$ for both parameters run from $0$ to some $I_{MAX}$. Here we are considering that $I_{MAX}=5i_{max}$ with $\alpha_{\mathrm{eff}[i>i_{max}]}=-10^6$. Hence, solving this linear system we obtain the profile $\alpha_{0[i]}$ on the galaxy plane. Such profile for our lattice disk galaxy model is plotted in figure~\ref{fig.alpha_0}.

Known the discrete reference profile for $\alpha_0$ and some symmetry along the orthogonal direction to the galaxy plane it is straight forward to compute the value of the functional parameter $\alpha_{\mathrm{eff}}$ outside the galaxy plan. Specifically at a spatial point orthogonal to the lattice position $[i,0]$ with a angle $\theta$ with respect to the $z$ axis, hence corresponding to the spherical coordinates $(r_{1[i]}/\sin(\theta),0,\theta)$, we obtain
\begin{equation}
\bar{\alpha}_{\mathrm{eff}[i]}(\theta)=-1+\frac{\displaystyle\sum_{[j,k]}\frac{M_{[j]}}{M_0}\left(\bar{\alpha}_{0[i]}(\theta)+1\right)\log\left(1-\bar{U}_{n[i,0][j,k]}(\theta)\right)}{\log\left(1-\bar{U}_{N[i]}(\theta)\right)}\ ,
\label{n_eff_sys}
\end{equation}
where $\bar{U}_{n[i,0][j,k]}(\theta)$ is the gravitational potential of the massive body at the lattice position $[j,k]$ and
$\bar{U}_{N[i]}(\theta)$ is the total gravitational potential of the $N$ bodies, both evaluated at the position $(r_{1[i]}/\sin(\theta),0,\theta)$
\begin{equation}
\begin{array}{rcl}
\bar{U}_{n[i,0][j,k]}(\theta)&=&\displaystyle\frac{2GM_{[j]}}{c^2\,\Delta \bar{r}_{1[i,0][j,k]}(\theta)}\ ,\\[5mm]
\bar{U}_{N[i]}(\theta)&=&\displaystyle \sum_{[j,k]}\bar{U}_{n[i,0][j,k]}(\theta)\ ,\\[5mm]
\Delta \bar{r}_{1[i,0][j,k]}(\theta)&=&\displaystyle \sqrt{\left(\frac{r_{1[i]}}{\sin\theta}\right)^2
-2r_{1[i]}r_{1[j]}\cos(\varphi_{[k]}) + r_{1[j]}^2}\ .
\end{array}
\end{equation}
In this expressions $\bar{r}_{1[i,0][j,k]}(\theta)$ is the Euclidean distance between the lattice point $[j,k]$ (i.e. $(r_{1[j]},\varphi_{[k]},\theta)$) and the spatial position $(r_{1[i]}/\sin(\theta),0,\theta)$.

As for $\bar{\alpha}_{0[i]}(\theta)$ it is the estimate for the value of the reference parameter $\alpha_0$ evaluated outside the galaxy plane
at the position $(r_{1[i]}/\sin(\theta),0,\theta)$. To actually estimate the value of this parameter at a generic $3D$ position it is required
to assume either spherical symmetry of each functional parameter $\alpha_n$ with respect to the center of mass of body $n$ or assuming an explicit anisotropy orthogonally to the galaxy plane proportional to $(\sin\theta)^{\xi_\theta}$ such that circular symmetry (for a fixed value of the coordinate $\theta$) is explicitly maintained. Such anisotropy is consistent with the planar model for the galaxy as the matter density distribution (a planar disk) explicitly violates spherical symmetry maintaining circular symmetry and it allows for a fine-tune of the mass-energy density maintaining it strictly positive outside the galaxy plane. In addition we note that as already discussed in the introduction, for large radial distances from the galaxy, the many body metric must consistently be asymptotically described by one single massive body with mass matching the total baryonic mass of the galaxy such that spherical symmetry is approximately recovered and the functional parameter $\alpha_{\mathrm{eff}}$ is asymptotically independent of both the angular coordinates $\varphi$ and $\theta$. Hence we are considering the following assumptions:
\begin{itemize}
\item approximately circular symmetry such that all quantities are independent of the angular coordinate $\varphi$;
\item anisotropic dependence of the parameter $\alpha_0$ on the factor $(\sin\theta)^{\xi_\theta}$ which, for the particular case of $\xi_\theta=0$, corresponds to spherical symmetry;
\item for large radial distances the parameter $\alpha_0$ is independent of the angular coordinate $\theta$ approximately matching the value $\alpha_{0[I_{MAX}]}$.
\end{itemize}
Given these assumptions it is straight forward to define the parameter $\alpha_0$ evaluated outside the galaxy plane at the position $(r_{1[i]}/\sin(\theta),0,\theta)$
\begin{equation}
\bar{\alpha}_{0[i]}(\theta)=\alpha_{0[I_{MAX}]}+\left(\tilde{\bar{\alpha}}_{0[i,0][j,k]}(\theta)-\alpha_{0[I_{MAX}]}\right)\left(\frac{\Delta \bar{r}_{1[i,0][j,k]}(\pi/2)}{\Delta \bar{r}_{1[i,0][j,k]}(\theta)}\right)^{\xi_\theta}\ ,
\end{equation}
where the sine of the angle $\theta_{[i,0][j,k]}(\theta)$ between the orthogonal direction to the lattice point $[j,k]$ and the line between
this lattice point and the position $(r_{1[i]}/\sin(\theta),0,\theta)$ is given by the ratio of the radial distances $\Delta \bar{r}_{1[i,0][j,k]}(\pi/2)/\Delta \bar{r}_{1[i,0][j,k]}(\theta)$ and $\tilde{\bar{\alpha}}_{0[i,0][j,k]}(\theta)$ is the value of the linearly interpolated reference parameter $\alpha_0$ corresponding to the distance to the origin of $\Delta \bar{r}_{1[i,0][j,k]}(\theta)$,
i.e. the distance to $\alpha_{0[0]}$~(\ref{alpha_0_int}).

\begin{figure}[!htbp]
\begin{center}
\includegraphics[width=120mm]{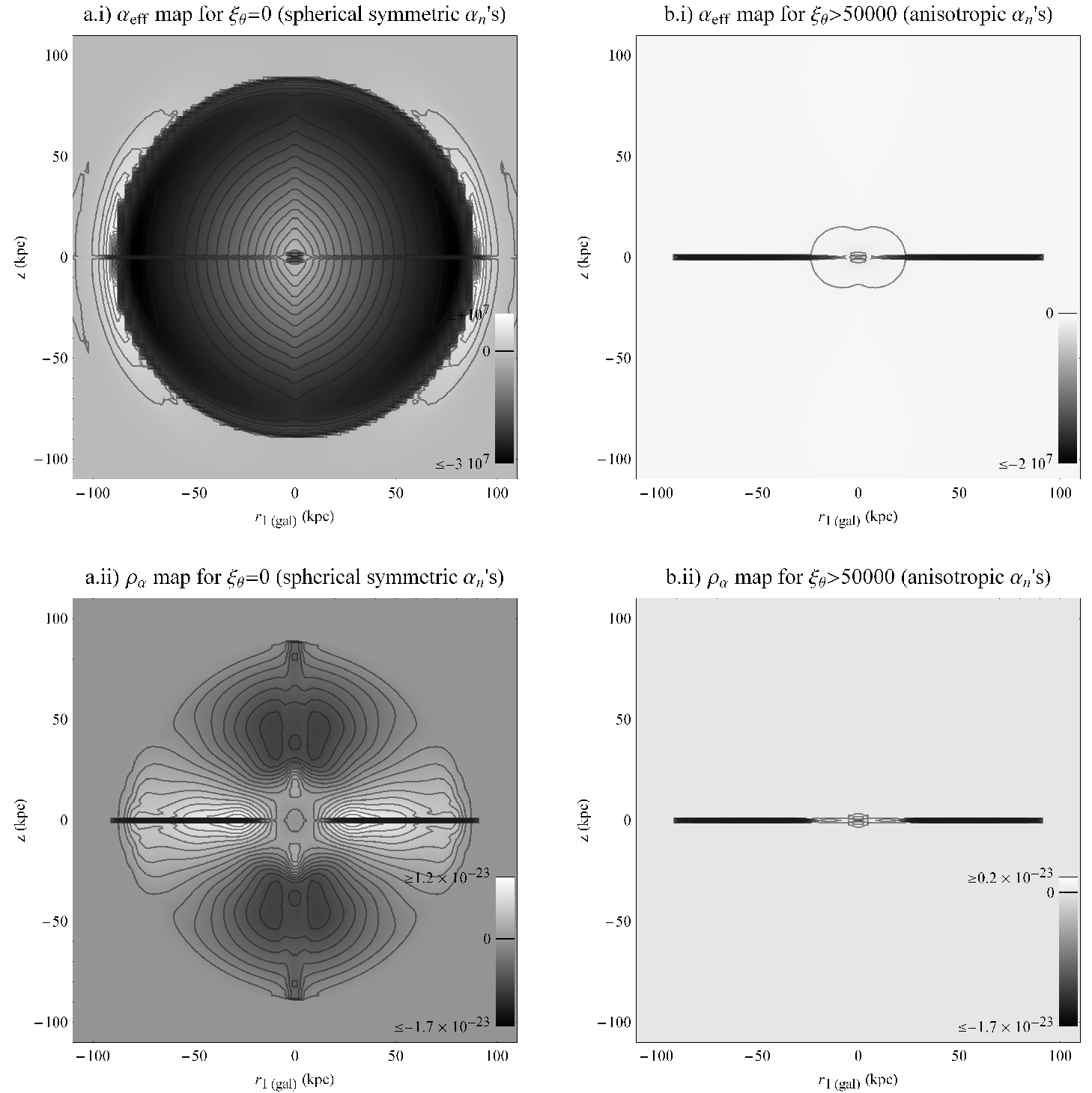}
\caption{\small The map of the functional parameter $\alpha_{\mathrm{eff}}$ and the mass-energy density $\rho_\alpha$ on a plane orthogonal to the galaxy containing the center of the galaxy, the $x$ axis corresponds to the galactic plane such that $r_{1(gal)}$ is the distance on the galaxy plane to the center of the galaxy and the $y$ axis corresponds to the orthogonal distance to the plane of the galaxy: {\bf a)} assuming spherical symmetry, $\xi_\theta=0$, for all $\alpha_n$'s there exist regions outside the galaxy plane where the mass-energy density is negative; {\bf b)} assuming anisotropy along the orthogonal direction to the galaxy plane, for $\xi_\theta> 50000$, the mass-energy density is strictly positive outside the galaxy plane.\label{fig.rho_z_map}}
\end{center}
\end{figure}

For $\xi_\theta=0$, corresponding to spherical symmetric $\alpha_n$'s there exist regions outside the galaxy plane where the mass-energy density is negative, while for large values of this exponent,
$\xi_\theta > 50000$, corresponding to anisotropic $\alpha_n$'s, the mass-energy density is strictly positive everywhere. The map for both the functional parameter $\alpha_{\mathrm{eff}}$ and the mass-energy density contribution $\rho_\alpha$ on a plane orthogonal to the galaxy containing the galaxy center is plotted in figure~\ref{fig.rho_z_map}. The several samplings of $\alpha_{\mathrm{eff}}$ and $\rho_\alpha$ orthogonal to the galaxy plane up to the distance (on the galaxy plane) to the center of the galaxy of $r_{1(gal)}=r_{1[61]}=73.48\,kpc$ considered when building the map of figure~\ref{fig.rho_z_map} are plotted in figures~\ref{fig.alpha_z_sph} to~\ref{fig.rho_z_ani} in the appendix.
\clearpage

Noting that, independently of the direction, for large radial distances to the center of the galaxy the functional parameter is approximately constant $\alpha_{\mathrm{eff}}=-10^6$ up to the upper cut-off $R_\alpha$~(\ref{alpha_big}), it is straight forward to estimate the relative mass-energy density contribution for the relative cosmological mass-energy density corresponding to the galaxy model developed in this work. Let us consider a relative large distance from the galaxy such that the gravitational field of the galaxy is approximately given by one single point-like massive body, hence the mass contribution in addition to the expanding Universe background mass-energy density $\rho_{RW}$ is
\begin{equation}
M_\alpha=4\pi\int_{0}^{+\infty}r_1^2(\rho_{\alpha}-\rho_{RW})dr_1=\frac{H^2}{2\,G}\left(\left(1-U(R_\alpha)\right)^{\alpha_{\mathrm{eff}}(R_\alpha)}-1\right)R_\alpha^3\ .
\end{equation}
Assuming that all the baryonic matter has a similar contribution to the relative cosmological mass-energy density contribution $M_\alpha\approx 4.96\,M_b$~\cite{wmap} we obtain for our galaxy model~(\ref{M_b})
\begin{equation}
M_{tot}=13.68\times 10^{11} M_\odot\ \ ,\ \ M_\alpha=11.38\times 10^{11}\,M_\odot\ \ \Leftrightarrow\ \ R_\alpha=9574.61\,kpc
\end{equation}
Recalling that within the $\Lambda CDM$ cosmological model such contribution must be included in the relative mass-energy density of Dark Matter, the extended background described by the ELA metric up to $R_\alpha$ is interpreted as the heuristic Dark Matter halo commonly employed to describe Dark Matter effects within galaxies.

\section{Discussion and Outlook}

By fitting the expanding locally anisotropic metric functional parameter $\alpha_{\mathrm{eff}}$~(\ref{metric_ELA_eff}) we have fully
describe the flattening of the rotation curve for the galaxy UGC2885. Based on a lattice model we have estimate values of this functional
parameter of order $\sim -10^{7}$ within the galaxy disk simultaneously maintaining compatibility with the well established short-scale
gravitational laws (the Schwarzschild metric is approximately recovered at planetary scales) and cosmological scale gravitational laws
(the Universe expansion). Phenomenologically this is interpreted as a unaccounted gravitational interaction acting on baryonic
matter that reproduces the effects attributed to Cold Dark Matter, namely the flattening of galaxies rotation curves. Hence these results constitute a novel local parameterization of Dark Matter distribution consistent with both local and large scale physical laws of the Universe which allows through the specific mass-energy profile computed to further investigate the properties of Dark Matter.  

We have also analyzed the mass-energy density on the galaxy plane showing that although there are negative contributions from the
extended background described by the ELA metric the total mass-energy density is strictly positive within the galaxy disk. Outside the galaxy disk,
both on the plane of the galaxy and outside the galaxy plane, to ensure that the mass-energy density is strictly positive, it is considered
a lower value for the functional parameter (specifically $\sim -10^{-6}$). This construction implies that the metric functional parameter is larger in absolute value in the vicinity of baryonic matter, i.e. on the galaxy plane, than outside of the galaxy such that the gravitational corrections are localized in the galaxy disk. In addition the functional parameter is null above the radial upper cut-off $R_\alpha=9574.61\,kpc$ such that the ratio of $M_\alpha$, the mass contribution of the extended background in addition to the cosmological background, to the the galaxy baryonic matter $M_b$ matches the ratio of Cold Dark Matter to baryonic relative cosmological mass-energy densities
$\Omega_c/\Omega_b=M_\alpha/M_b=4.96$ consistently with the physical interpretation that the ELA metric constitutes a parameterization of Cold Dark Matter.

A relevant question raised by the construction developed here is whether the gravitational corrections due to the expanding locally anisotropic background to the redshift of radiation emitted from the galaxy are significant or not~\cite{Pioneer}. To answer this question let us recall that when computing the profile for the functional parameter we have excluded the point-like masses at each lattice point as these are considered test
masses on the background of the remaining $N-1$ massive bodies on the galaxy. When considering an external point mass in the vicinity of each
body on the galaxy, the effective functional parameter for the $N$ body system, as perceived by the external test mass, must decrease in absolute value as, near massive objects, the Schwarzschild metric is asymptotically recover. Hence for a external test mass in the galaxy plane, near each massive body the functional parameter decreases in absolute value 
such that the red-shift corrections are negligible for radiation emitted from each body on the galaxy. As an analogy let us note that for many-body gravitational systems on Schwarzschild geometries, when computing orbits a given planet is considered a test mass such that its own gravitational field is not accounted in the equations of motion, however the frequency-shift for radiation emitted from each planet is mainly accounted for by the planet gravitational field.

As future directions of research we note that to develop a more detailed model for galaxies parameterized
by a many body expanding locally anisotropic background it must be derived a continuous description of the intergalactic matter (gas) not contained on stars. Also, within a more detailed model, the assumption that all masses have a similar profile for the functional parameter given by the scaling law~(\ref{scaling_M}) may be an over simplification as the localization of the gravitational corrections in the vicinity of
the galactic plane indicates that the functional parameter, for each of the many bodies on the galaxy, may depend on the remaining masses
in the gravitational system, hence not being exactly circular symmetric as considered in the simplified model developed in this work.

As a final remark let us note that a direct analysis in the neighborhood of galaxies which may either confirm or dismiss the construction suggested here is achievable, for instance, by analyzing the lens effect of the background radiation near galaxies. Also within the Solar System
a test mass orbiting orthogonally to the planetary system plane would allow for a direct measurement of the localization
of the gravitational corrections encoded in the expanding locally anisotropic metric~\cite{AU}. We leave these analysis to another work.\\

\noindent{\sc\bf Acknowledgments}:
Work supported by Portuguese Foundation for Science and Technology (FCT) through grant SFRH/BPD/34566/2007 up to 2014 and by the Portuguese Foundation for Science and Technology (FCT) and European Union through the Centro2020 and MATIS (CENTRO-01-0145-FEDER-000014) from 2017 onwards.

\appendix

\section{Appendix}
\begin{figure}[!htbp]
\begin{center}
\includegraphics[width=120mm]{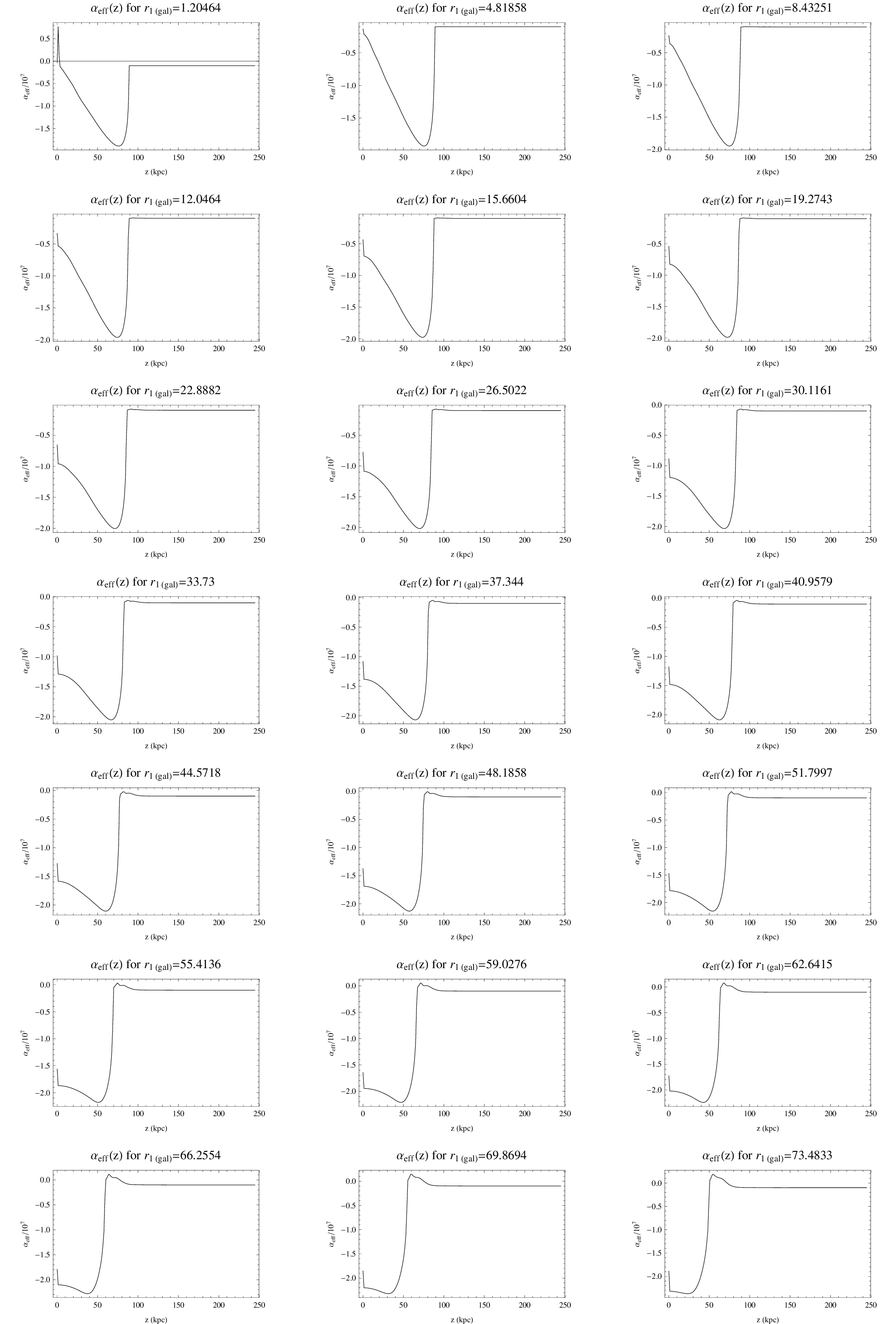}
\caption{\small Estimates for $\alpha_{\mathrm{eff}}$ outside of the galaxy plane assuming spherical symmetry of the functional parameters $\alpha_n$'s ($\xi_\theta=0$). \label{fig.alpha_z_sph}}
\end{center}
\end{figure}

\begin{figure}[!htbp]
\begin{center}
\includegraphics[width=120mm]{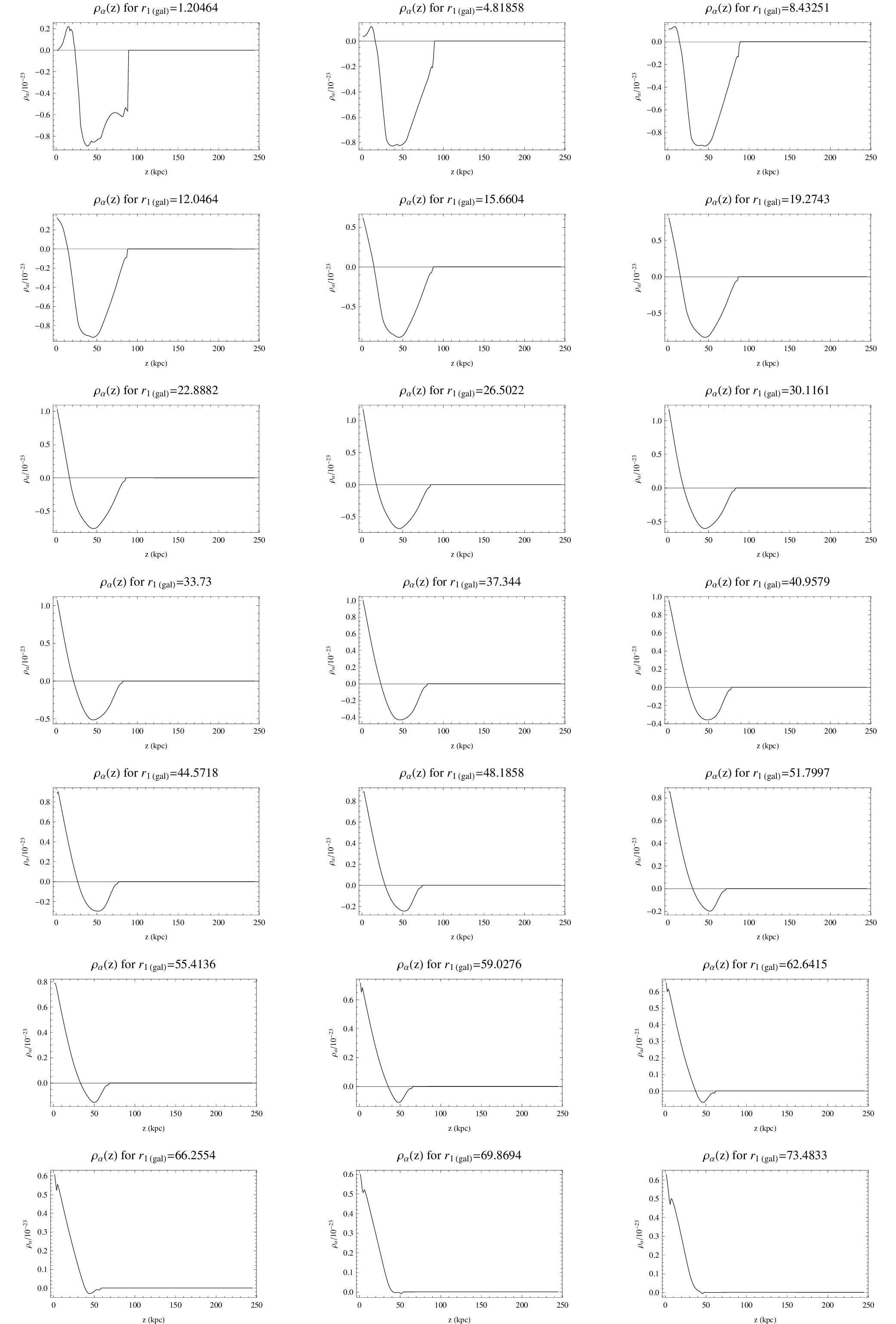}
\caption{\small Estimates for $\rho_\alpha$ outside of the galaxy plane assuming spherical symmetry of the functional parameters $\alpha_n$'s ($\xi_\theta=0$). \label{fig.rho_z_sph}}
\end{center}
\end{figure}

\begin{figure}[!htbp]
\begin{center}
\includegraphics[width=120mm]{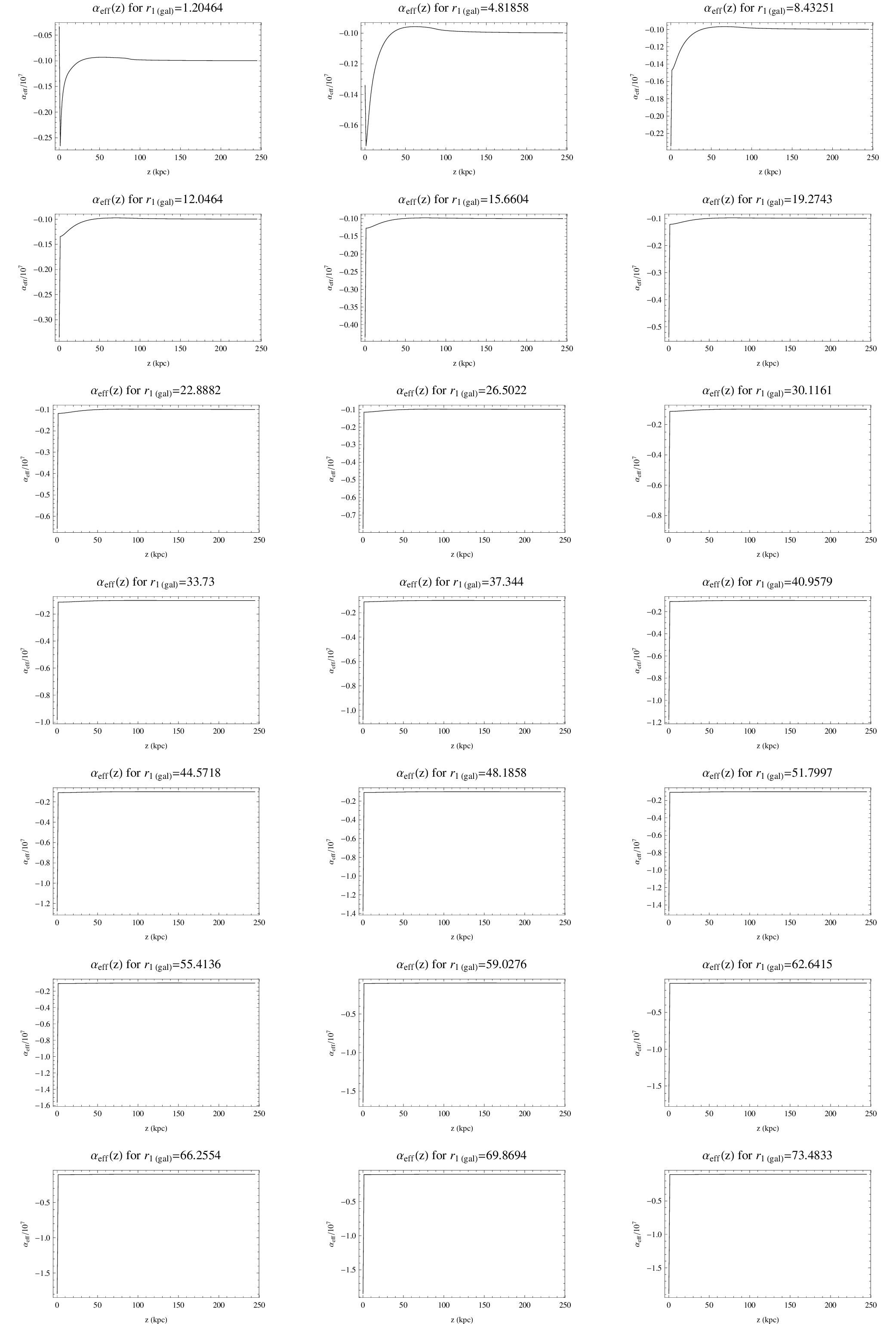}
\caption{\small Estimates for $\alpha_{\mathrm{eff}}$ outside of the galaxy plane assuming anisotropy of the functional parameters $\alpha_n$'s ($\xi_\theta> 50000$). \label{fig.alpha_z_ani}}
\end{center}
\end{figure}

\begin{figure}[!htbp]
\begin{center}
\includegraphics[width=120mm]{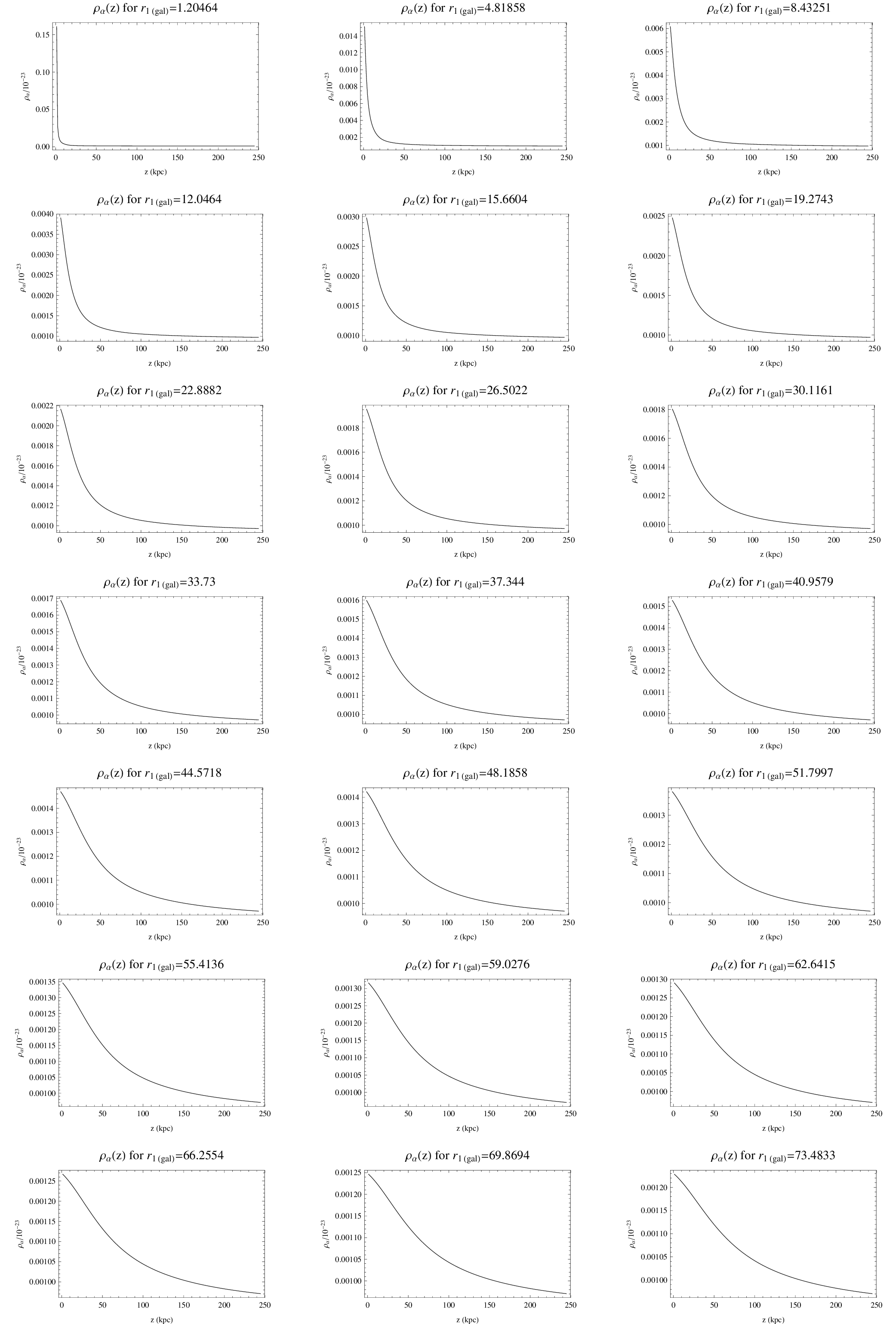}
\caption{\small Estimates for $\rho_\alpha$ outside of the galaxy plane assuming anisotropy of the functional parameters $\alpha_n$'s ($\xi_\theta> 50000$). \label{fig.rho_z_ani}}
\end{center}
\end{figure}

In this appendix we list the values for the galaxy model masses, functional parameter and velocity contributions corresponding to figure~\ref{fig.V} in table~\ref{table.V} and the samplings along the orthogonal to the galaxy plane of the functional parameter and mass-energy densities considered to plot the maps of figure~\ref{fig.rho_z_map} in figures~\ref{fig.alpha_z_sph} to~\ref{fig.rho_z_ani}.

\begin{table}[ht]
\begin{center}
{\tiny
\begin{tabular}{r|cccccccc}
$i$&$\frac{r_{1[i]}}{kpc}$&$\frac{M_{[i]}}{10^8M_\odot}$&$\frac{\alpha_{\mathrm{eff}[i]}}{10^7}$&$V_N(km\,s^{-1})$&$V_\alpha(km\,s^{-1})$&$V_{\mathrm{orb}}(km\,s^{-1})$\\\hline\hline
0&0      &75.3802& 0.0000003  &0&0&0\\
1&1.20464&4.48560& -0.0335008 &208.788  & -0.143264& 208.788\\
2&2.40929&4.73906& -0.067002  &210.623  & -0.434526& 210.623\\
3&3.61393&4.78543& -0.100503  &212.459  & -0.911588& 212.457\\
4&4.81858&5.05407& -0.134004  &214.294  & -1.612610& 214.288\\
5&6.02322&4.99917& -0.167506  &216.129  & -2.584880& 216.114\\
6&7.22787&5.22984& -0.201007  &217.965  & -3.848380& 217.931\\
7&8.43251&5.15292& -0.234508  &219.804  & -5.457760& 219.736\\
8&9.63715&5.38540& -0.268009  &221.648  & -7.377250& 221.525\\
9&10.8418&5.32973& -0.301511  &223.492  & -9.671340& 223.283\\
10&12.0464&5.63228& -0.335012 &226.963 & -12.06130& 226.642\\
11&13.2511&5.64015& -0.368513 &237.720 & -13.82750& 237.318\\
12&14.4557&5.58059& -0.402014 &248.478 & -14.62600& 248.047\\
13&15.6604&5.36167& -0.435516 &259.235 & -13.40730& 258.889\\
14&16.8650&5.05500& -0.469017 &266.086 & -9.44566& 265.918\\
15&18.0697&4.81708& -0.502518 &273.043 & 0.00000& 273.043\\
16&19.2743&4.41684& -0.538329 &279.600 & 10.7279& 279.805\\
17&20.4790&4.04947& -0.577473 &282.205 & 21.7037& 283.038\\
18&21.6836&3.67934& -0.617500 &282.686 & 32.7323& 284.575\\
19&22.8882&3.37378& -0.657534 &282.727 & 43.8787& 286.112\\
20&24.0929&3.02316& -0.697586 &281.168 & 54.9193& 286.481\\
21&25.2975&2.76502& -0.737376 &277.601 & 65.6203& 285.252\\
22&26.5022&2.50679& -0.776331 &273.602 & 76.2271& 284.022\\
23&27.7068&2.31885& -0.814293 &269.214 & 86.7563& 282.847\\
24&28.9115&2.12502& -0.850928 &264.624 & 97.2881& 281.941\\
25&30.1161&1.99283& -0.885900 &259.543 & 107.787& 281.035\\
26&31.3207&1.87052& -0.918503 &256.006 & 113.723& 280.129\\
27&32.5254&1.75714& -0.950966 &252.866 & 120.244& 280.000\\
28&33.7300&1.65179& -0.983126  &249.919  & 126.257& 280.000\\
29&34.9347&1.55366& -1.01515  &247.094  & 131.699& 280.000\\
30&36.1393&1.46206& -1.04723  &244.360  & 137.085& 280.186\\
31&37.3440&1.37638& -1.07935  &241.697  & 142.334& 280.493\\
32&38.5486&1.29612& -1.11154  &239.091  & 147.256& 280.800\\
33&39.7533&1.22087& -1.14388  &236.533  & 151.901& 281.108\\
34&40.9579&1.15027& -1.17643  &234.013  & 156.309& 281.415\\
35&42.1625&1.08400& -1.20928  &231.526  & 160.510& 281.723\\
36&43.3672&1.02181& -1.24248  &229.067  & 164.732& 282.150\\
37&44.5718&0.963453& -1.27602  &226.635 & 170.630& 283.686\\
38&45.7765&0.908717& -1.30932 &224.227 & 176.280& 285.223\\
39&46.9811&0.857398& -1.34237 &221.843 & 181.706& 286.760\\
40&48.1858&0.809309& -1.37511 &219.484 & 186.893& 288.275\\
41&49.3904&0.764268& -1.40752 &217.151 & 191.888& 289.785\\
42&50.5951&0.722103& -1.43954 &214.845 & 196.710& 291.295\\
43&51.7997&0.682648& -1.47110 &212.568 & 201.629& 292.983\\
44&53.0043&0.645745& -1.50203 &210.322 & 206.611& 294.828\\
45&54.2090&0.611239& -1.53215 &208.109 & 211.435& 296.672\\
46&55.4136&0.578984& -1.56131 &205.931 & 215.399& 298.000\\
47&56.6183&0.548839& -1.58957 &203.789 & 217.426& 298.000\\
48&57.8229&0.520670& -1.61739 &201.686 & 219.378& 298.000\\
49&59.0276&0.494349& -1.64469 &199.622 & 221.257& 298.000\\
50&60.2322&0.469754& -1.67139 &197.601 & 223.065& 298.000\\
51&61.4369&0.446771& -1.69740 &195.621 & 224.803& 298.000\\
52&62.6415&0.425290& -1.72261 &193.686 & 226.472& 298.000\\
53&63.8461&0.405209& -1.74692 &191.797 & 228.074& 298.000\\
54&65.0508&0.386431& -1.77019 &189.955 & 229.611& 298.000\\
55&66.2554&0.368865& -1.79227 &188.161 & 231.084& 298.000\\
56&67.4601&0.352427& -1.81298 &186.417 & 232.493& 298.000\\
57&68.6647&0.337035& -1.83213 &184.726 & 233.839& 298.000\\
58&69.8694&0.322617& -1.84947 &183.090 & 235.122& 298.000\\
59&71.0740&0.309100& -1.86471&181.513 & 236.341& 298.000\\
60&72.2787&0.296422& -1.87751 &180.000 & 237.495& 298.000\\
61&73.4833&0.284521& -1.88744 &178.559 & 238.580& 298.000\\
62&74.6879&0.273341& -1.89396 &177.202 & 239.590& 298.000\\
63&75.8926&0.262829& -1.89639 &175.945 & 240.515& 298.000\\
64&77.0972&0.252937& -1.89387 &174.816 & 241.337& 298.000\\
65&78.3019&0.243619& -1.88521 &173.863 & 242.024& 298.000\\
66&79.5065&0.234833& -1.86884 &173.175 & 242.517& 298.000\\
67&80.7112&0.226541& -1.84251 &172.945 & 242.681& 298.000\\
68&81.9158&0.218707& -1.80296 &173.690 & 242.149& 298.000\\
69&83.1205&0.211297& -1.74546 &177.812 & 239.138& 298.000\\
70&84.3251&0.186052& -1.66494 &185.471 & 233.248& 298.000\\
71&85.5297&0.132826& -1.55204 &189.782 & 227.469& 296.242\\
72&86.7344&0.0824872& -1.37634 &188.435 & 226.289& 294.473\\
73&87.9390&0.0352501& -1.04070 &183.142 & 228.318& 292.694\\
74&89.1437&0.00180299& -0.10000 &172.674 & -3.87951& 172.63\\
75&90.3483&0        & -0.10000 &163.020 & -4.14655& 162.967\\

\hline
\end{tabular}
}
\caption{\small Values of the point-like masses $M_{[i]}$, functional parameter $\alpha_{\mathrm{eff}[i]}$,
and velocity contributions $V_{N[i]}$ and $V_{\alpha[i]}$ to $V_{\mathrm{orb}[i]}$. \label{table.V}}
\end{center}
\end{table}

\end{document}